\documentstyle[12pt,epsfig,epstopdf,subfigure,color,amsfonts,amssymb,inputenc,graphicx,amsmath,amsxtra,amstext,slashed]{article}
\newcommand{\beq}{\begin{equation}}
\newcommand{\eeq}{\end{equation}}
\newcommand{\bea}{\begin{eqnarray}}
\newcommand{\eea}{\end{eqnarray}}
\newcommand{\al}{\alpha}

\newcommand{\vep}{\varepsilon}
\newcommand{\ep}{\epsilon}

\newcommand{\der}{\partial}

\newcommand{\m}{\mu}
\newcommand{\n}{\nu}

\newcommand{\nn}{\nonumber}

\begin{document}

\baselineskip=15.5pt \pagestyle{plain} \setcounter{page}{1}
%
\begin{titlepage}

\vskip 0.8cm

\begin{center}

{\Large \bf  Deep inelastic scattering from polarized spin-$1/2$
hadrons at low $x$ from string theory}

\vskip 1.cm

{\large {{\bf Nicolas Kovensky}{\footnote{\tt
nico.koven@fisica.unlp.edu.ar}}, {\bf Gustavo Michalski}{\footnote{\tt
michalski@fisica.unlp.edu.ar}}, {\bf and Martin
Schvellinger}{\footnote{\tt martin@fisica.unlp.edu.ar}}}}

\vskip 1.cm

{\it Instituto de F\'{\i}sica La Plata-UNLP-CONICET. \\
Boulevard 113 e 63 y 64, (1900) La Plata, Buenos Aires, Argentina. \\
and \\
Departamento de F\'{\i}sica, Facultad de Ciencias Exactas,
Universidad Nacional de La Plata. \\
Calle 49 y 115, C.C. 67, (1900) La Plata, Buenos Aires, Argentina.} \\

\vspace{1.cm}

{\bf Abstract}

\vspace{1.cm}

\end{center}

We study polarized deep inelastic scattering of charged leptons from
spin-$1/2$ hadrons at low values of the Bjorken parameter and large
't Hooft coupling in terms of the gauge/string theory duality. We
calculate the structure functions from type IIB superstring theory
scattering amplitudes. We discuss the role of the non-Abelian
Chern-Simons term and the Pauli term from the five-dimensional
$SU(4)$ gauged supergravity. Furthermore, the exponentially
small-$x$ regime where Regge physics becomes important is analyzed
in detail for the antisymmetric structure functions. In this case
the holographic dual picture of the Pomeron exchange is realized by
a Reggeized gauge field. We compare our results with experimental
data of the proton antisymmetric structure function $g_1$, obtaining
a very good level of agreement.

\noindent

\end{titlepage}

\newpage

{\small \tableofcontents}

\newpage

\section{Introduction}

The idea of this work is to study polarized deep inelastic
scattering (DIS) of charged leptons off spin-$1/2$ hadrons, in order
to investigate properties of the hadronic tensor at small values of
the Bjorken parameter. We consider large values of the 't Hooft
coupling $\lambda$ and the planar limit of the gauge theory, within
the framework of the gauge/string theory duality. We carry out first
principles calculations starting from type IIB superstring theory
scattering amplitudes. Alternatively, we show how to approach the
problem by deriving heuristic Lagrangians for the symmetric and the
antisymmetric contributions. We first introduce the heuristic
approach which is more intuitive, and then we describe the formal
string theoretical derivation. The parametric region we focus on is
$x \ll 1/\sqrt{\lambda}$, where type IIB supergravity does not give
an accurate description of the holographic dual DIS process, hence
it is necessary to consider string theory. Furthermore, we
investigate the region where the Bjorken parameter becomes
exponentially small, which allows us to compare our results for the
antisymmetric structure function $g_1$ with recent experimental data
of electron-proton DIS.

The DIS differential cross-section of a charged lepton off a hadron
is proportional to the contraction of the leptonic tensor, which is
obtained from perturbative QED, and the hadronic tensor, where
non-perturbative QCD effects are essential. The hadronic tensor of a
spin-$1/2$ hadron is usually written in terms of symmetric (S) and
antisymmetric (A) tensors under Lorentz indices exchange
\cite{Anselmino:1994gn,Lampe:1998eu}\footnote{We use the notation
for the hadronic tensor as in reference \cite{Gao:2010qk}, having
some sign differences with respect to
\cite{Anselmino:1994gn,Lampe:1998eu}.}
\bea
W_{\mu\nu}&=&W^{\mathrm{(S)}}_{\mu\nu}(q,P)+i \,
W^{\mathrm{(A)}}_{\mu\nu}(q,P,S) \, , \nn\\
&& \nn \\
W^{\mathrm{(S)}}_{\mu\nu}&=& \left(\eta_{\mu\nu}-\frac{q_\mu
q_\nu}{q^2}\right) \left[F_1(x,q^2)+\frac{1}{2} \frac{S\cdot
q}{P\cdot q}g_5(x,q^2)\right] \, , \nn \\
&& -\frac{1}{P\cdot q} \left(P_\mu - \frac{P\cdot q}{q^2} q_\mu
\right) \left(P_\nu - \frac{P\cdot q}{q^2} q_\nu \right)
\left[F_2(x,q^2)+\frac{S\cdot q}{P\cdot q}g_4(x,q^2)\right] \nn\\
&&
-\frac{1}{2 P\cdot q}\left[\left(P_\mu - \frac{P\cdot q}{q^2} q_\mu
\right) \left(S_\nu - \frac{S\cdot q}{P\cdot q} P_\nu \right)+
\left(P_\nu - \frac{P\cdot q}{q^2} q_\nu \right)\left(S_\mu -
\frac{S\cdot q}{P\cdot q} P_\mu \right)\right] \nn \\
&& g_3(x,q^2) \, , \nn \\
W^{\mathrm{(A)}}_{\m\n}&=&-\frac{\vep_{\mu\nu\rho\sigma}q^\rho}{P\cdot
q} \left\{S^\sigma g_1(x,q^2) + \left[S^\sigma-\frac{S\cdot q}
{P\cdot q}P^\sigma\right]g_2(x,q^2)\right\}
-\frac{\vep_{\mu\nu\rho\sigma}q^\rho P^\sigma}{2P\cdot q} F_3(x,q^2)
\, , \nn \\
&&
\label{w}
\eea
where $\eta_{\mu\nu}=\mathrm{diag}(-1,1,1,1)$, $P_\mu$ and $S_\mu$
are the four-momentum and the spin vector of the incident hadron,
respectively. The four-momentum of the virtual photon is denoted by
$q_\mu$. The symmetric structure functions are $F_1$, $F_2$, $g_3$,
$g_4$ and $g_5$, while $F_3$, $g_1$ and $g_2$ are the antisymmetric
ones. For electromagnetic DIS in QCD the non-preserving parity
structure functions $g_3$, $g_4$, $g_5$ and $F_3$ vanish. In fact,
we consider electromagnetic DIS not precisely for QCD but for an IR
deformation of ${\cal {N}}=4$ SYM theory. The last is a chiral
theory, therefore it may lead to a non-vanishing $F_3$. In this
sense this result is in perfect agreement with respect to the
glueball case presented in reference \cite{Kovensky:2017oqs}. The
condition for $F_3$ to be non-vanishing is that the IR deformation
of ${\cal {N}}=4$ SYM theory must be such that there are massless
Nambu-Goldstone modes associated to the spontaneously broken
$R$-symmetry \cite{Hatta:2009ra}. We will assume this property in
the present approach.

The Bjorken parameter is defined as
\begin{equation}
x=-\frac{q^2}{2P\cdot q} \, ,
\end{equation}
being the physical range $0 \leq x \leq 1$, in the DIS limit $q^2
\gg P^2$ while $x$ is kept fixed. From the Cutkosky rules for
scattering amplitudes, based on S-matrix theory, one can derive the
optical theorem leading to the following relations
\begin{equation}
W^{\m\n}_{({\mathrm{S}})} = 2 \pi
{\mathrm{Im}}\left[T^{\m\n}_{({\mathrm{S}})}\right]
\,\,\,\,\,\,\,\,\,\,\ , \ W^{\m\n}_{({\mathrm{A}})} = 2 \pi
{\mathrm{Im}}\left[T^{\m\n}_{({\mathrm{A}})}\right] \, ,
\label{relWT}
\end{equation}
where the tensor $T^{\m\n}$ is defined by the time-ordered
expectation value of two electromagnetic currents inside the hadron
\begin{equation}
T^{\m\n}\equiv i \int d^{4}x e^{i q\cdot x} \langle P|
{\mathrm{\hat{T}}} \{J^\mu(x) J^\nu(0)\} |P \rangle \, .
\end{equation}
This relates DIS to forward Compton scattering (FCS), which is what
one calculates. DIS and FCS are schematically shown in figure 1.

\begin{figure}[ht]
\centering
\begin{subfigure}[]{
\includegraphics[scale=0.35]{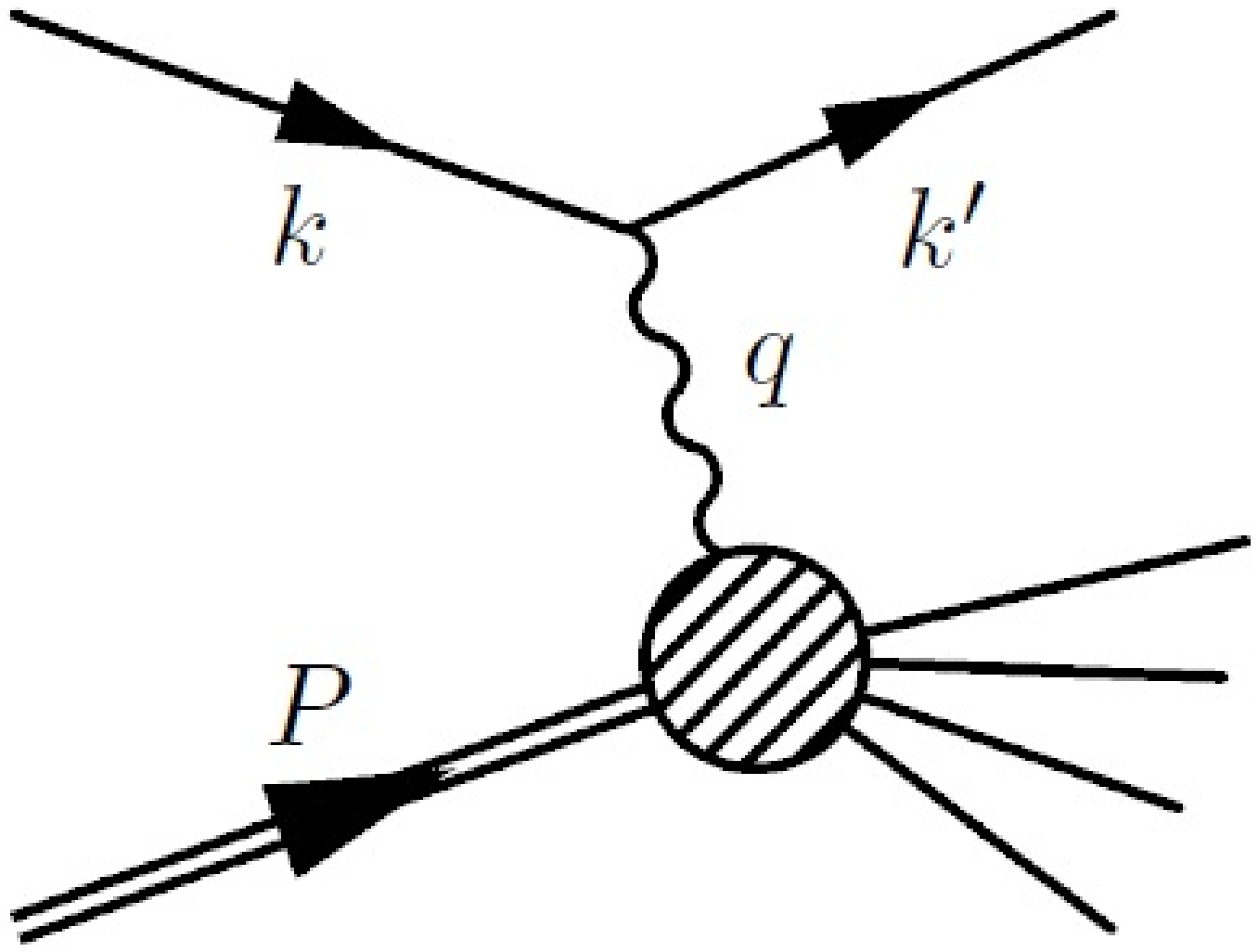}}
\end{subfigure}
\begin{subfigure}[]{
\includegraphics[scale=0.2]{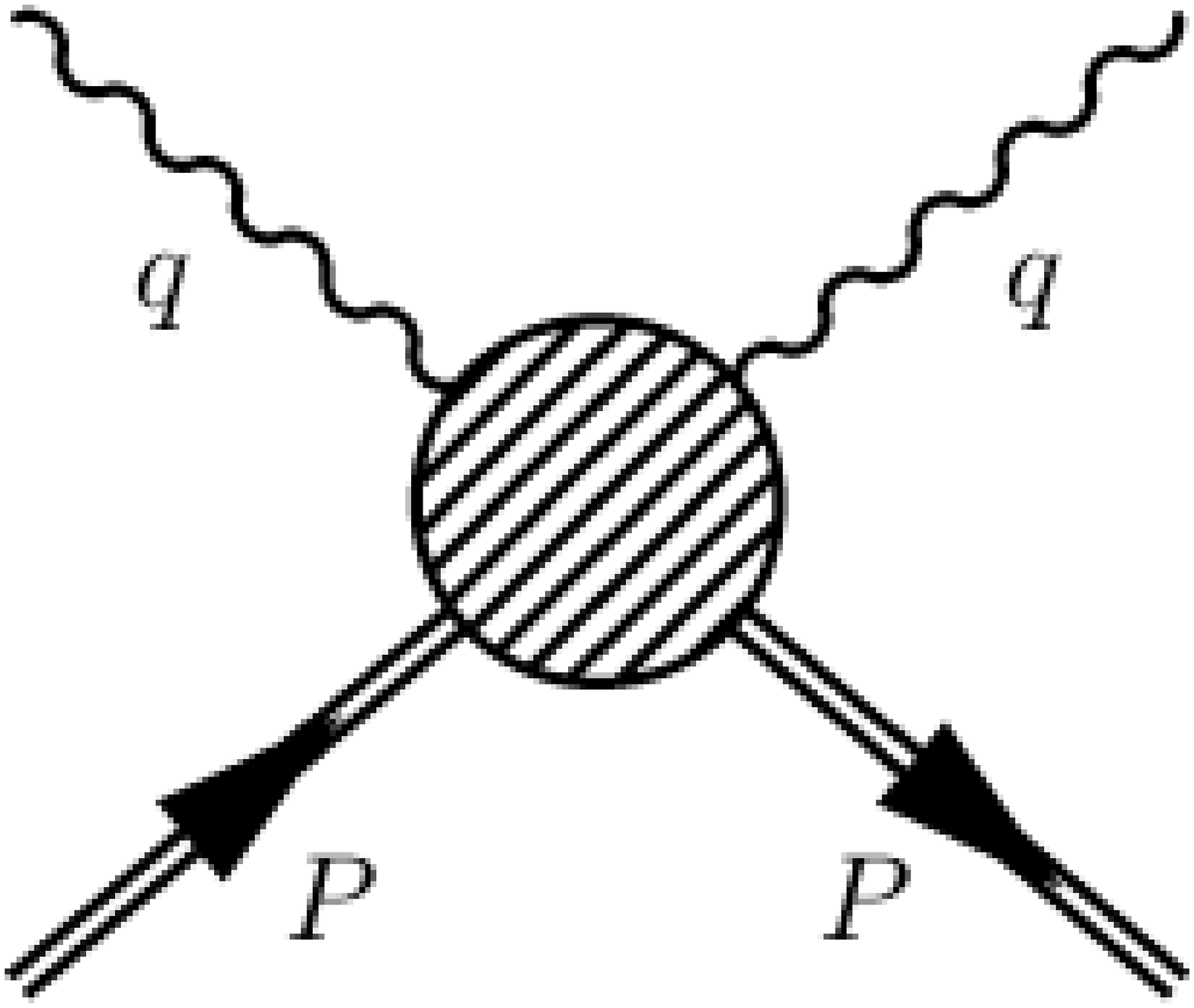}}
\end{subfigure}
\label{DISFCS} \caption{\small Schematic representation of DIS (a)
and FCS (b) processes. $k$ and $k'$ denote the four-momenta of the
incoming and outgoing leptons in DIS.}
\end{figure}

In the pioneering work by Polchinski and Strassler
\cite{Polchinski:2002jw}, the symmetric structure functions $F_1$
and $F_2$ of DIS of a charged lepton off a spin-$1/2$ hadron have
been calculated in the supergravity regime, {\it i.e.} for
$1/\sqrt{\lambda} \ll x < 1$. The spin-$1/2$ hadron can be
holographically represented by a dilatino wave-function in the bulk
of AdS$_5 \times S^5$ in type IIB supergravity, with the inclusion
of an IR cut-off $z_0=1/\Lambda$. This scale breaks conformal
invariance in the IR of the holographic dual gauge field theory,
inducing color confinement. The holographic dual gauge theory
corresponds to the planar limit of ${\cal {N}}=4$ $SU(N)$
supersymmetric Yang-Mills (SYM) theory in four dimensions, with an
IR cut-off scale $\Lambda$. In the UV this gauge theory is
conformal. The holographic dual process is schematically represented
in figure 2. The result from \cite{Polchinski:2002jw} for the
symmetric structure functions is
\beq
2 F_1 = F_2 =  \pi \, A' \, {\cal {Q}}^2 \,
\left(\frac{\Lambda^2}{q^2}\right)^{\tau-1} \, x ^{\tau+1} \,
(1-x)^{\tau-2} \, , \label{PSdilatino}
\eeq
where $A'$ is a dimensionless constant, ${\cal {Q}}$ is a charge
eigenvalue under the $U(1) \subset SU(4)$ symmetry group, and $\tau$
is the twist of the incident hadron, $\tau=\Delta-s$, being $\Delta$
the conformal dimension and $s$ the spin (in the present case
$s=1/2$).

\begin{figure}[ht]
\centering
\includegraphics[scale=0.4]{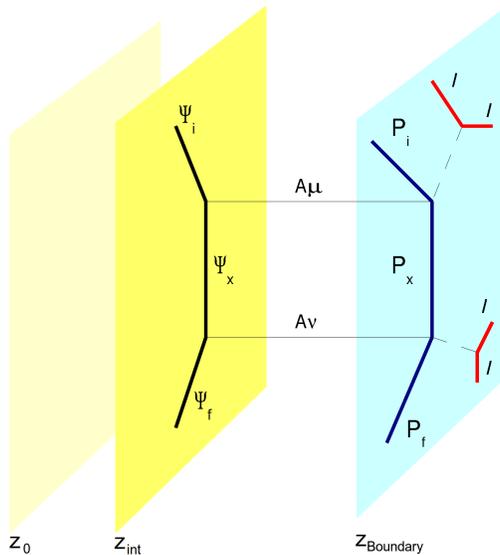}
\caption{\small Schematic picture of the $s$-channel diagram
corresponding to the holographic dual description of forward Compton
scattering in the $1/\sqrt\lambda \ll x < 1$ regime. The incoming
and outgoing spin-$1/2$ hadrons with four-momenta $P_i$ and $P_f$
are represented by blue lines in the boundary theory. Their
corresponding dual dilatino fields in the bulk are denoted by
$\Psi_i$ and $\Psi_f$, respectively. Gauge fields $A_\mu$ and
$A_\nu$ couple to the $J^\mu(x)$ and $J^\nu(0)$ electromagnetic
currents in the boundary gauge field theory. $z_0$ is the IR cut-off
and $z_{int}$ is where the graviphoton-dilatino interaction takes
place. Red lines denote leptons ($l$), while dashed lines indicate
virtual photons.}
\end{figure}

Furthermore, also in the supergravity regime ($1/\sqrt{\lambda} \ll
x < 1$), in reference \cite{Gao:2009ze} polarized DIS structure
functions considering a spin-$1/2$ hadron have been studied using
the AdS/CFT duality. The results are \cite{Gao:2009ze}
\beq
2 F_1 = F_2 = F_3 = 2 g_1 = g_i \, , \,\,\,\,\,\,\,
g_2 = \left( \frac{1}{2 x}
\frac{\tau+1}{\tau-1}-\frac{\tau}{\tau-1}\right) \, g_1 \, ,
\eeq
where $i=3, 4, 5$. The explicit form of $F_2$ is given in equation
(\ref{PSdilatino}). The functions $F_3$, $g_3$, $g_4$ and $g_5$ are
similar to $F_2$ since the dilatino is a right-handed fermion in the
massless limit. However, as we shall show in section 2, $g_3$, $g_4$
and $g_5$ vanish at leading order in $1/N$ for $x \ll
1/\sqrt{\lambda}$. Further calculations in this regime for
non-forward Compton scattering have been done in \cite{Gao:2009se}.
A study of neutral spin-$1/2$ hadrons (similarly to case of charged
spin-$1/2$ hadrons considered in \cite{Gao:2009ze}) is presented in
reference \cite{Gao:2010qk} for this regime of the Bjorken
parameter.

\begin{figure}[ht]
\centering
\includegraphics[scale=0.4]{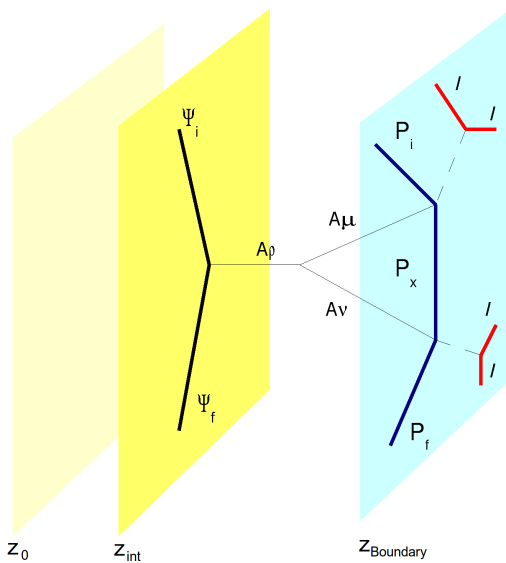}
\caption{\small
Schematic picture of the $t$-channel diagram corresponding to the
holographic dual description of forward Compton scattering in the $x
\ll 1/\sqrt\lambda$ regime. The incoming and outgoing spin-$1/2$
hadrons with four-momenta $P_i$ and $P_f$ are indicated with blue
lines in the boundary theory. Their corresponding dual fields in the
bulk are denoted by $\Psi_i$ and $\Psi_f$, respectively. Gauge
fields $A_\mu$ and $A_\nu$ couple to the $J^\mu(x)$ and $J^\nu(0)$
electromagnetic currents in the boundary gauge field theory.}
\end{figure}
~

On the other hand, in a completely different physical regime as it
is the exponentially small-$x$ region, the proton $F_2$ structure
function has been investigated by Brower and collaborators in
\cite{Brower:2010wf}, by using the BPST-Pomeron techniques developed
by Brower, Polchinski, Strassler and Tan within the gauge/string
theory duality framework \cite{Brower:2006ea}. The authors of
reference \cite{Brower:2010wf} have found that the BPST kernel fits
remarkably well the region where the four-momentum transfer $q^2$ of
the virtual photon is large, and also it works surprisingly well for
small values of $q^2$, as low as $q^2=0.1$ (GeV/c)$^2$. They fit
their result for $F_2$ to the combined H1-ZEUS small-$x$ data of the
inclusive DIS cross sections measured by H1 and ZEUS Collaborations
in neutral and charged current unpolarized $e^\pm \, p$ scattering
at HERA \cite{Aaron:2009aa,Breitweg:1998dz,Chekanov:2001qu}, in the
range $0.1$ (GeV/c)$^2 \leq q^2 \leq 400$ (GeV/c)$^2$, and for
$10^{-6} \leq x \leq 10^{-2}$. For large $q^2$ conformal symmetry
dominates, while near to the IR the hard-wall cut-off becomes
important. This behavior is reflected on the results presented in
\cite{Brower:2006ea}. In addition, in the case of ${\cal {N}}=4$
$SU(N)$ SYM theory considering polarized DIS also from a spin-$1/2$
hadron, a heuristic calculation based on the AdS/CFT duality has
been developed in \cite{Hatta:2009ra}. The result is that the
Reggeized virtual photon leads to the polarized structure functions
$F_3$ and $g_1$. For exponentially small $x$ it has been obtained
that $g_1 \approx (1/x)^{1-1/(2 \sqrt{\lambda})}$.

In the present work we derive explicitly all the structure functions
for a spin-$1/2$ hadron in the low-$x$ regime. For small but not
exponentially small $x$, in addition to a heuristic derivation, we
carry out a detailed top-down string theory calculation from closed
strings scattering amplitudes which constitutes the first complete
derivation of this kind for spin-$1/2$ hadron in the low-$x$ regime.
This approach leads to effective Lagrangians from which one can
construct the leading-diagram contributions which are $t$-channel
Feynman diagrams in the bulk theory. Their corresponding schematic
representation is shown in figure 3. This is related to the Feynman
diagrams presented in figure 4, corresponding to the calculations of
the symmetric and antisymmetric structure functions in the range
$\exp{(-\sqrt\lambda)} \ll x \ll 1/\sqrt\lambda$ that we introduce
in sections 2 and 3, respectively.

Furthermore, for the exponentially low-$x$ regime, generalizing the
BPST-Pomeron approach, we consider a Reggeized gauge field and
derive the antisymmetric structure functions. Then, we compare with
experimental data. We fit our results\footnote{As we shall explain
in sections 4 and 5 we consider two different models, namely: a
conformal model with no IR cut-off and the hard-wall model that we
have already described.} of $g_1$ to the data of the corresponding
structure function of the proton at $190$ GeV measured by the SMC
Collaboration \cite{Adeva:1998vv}, and by the COMPASS Collaboration
with beam energies of $160$ GeV and $200$ GeV reported in
\cite{Alekseev:2010hc} and \cite{Adolph:2015saz}, respectively. In
these cases we consider data within the $x<0.01$ region. Following
\cite{Adolph:2015saz}, in our figures 6 and 7 we also include data
from the SMC \cite{Adeva:1998vv}, EMC \cite{Ashman:1987hv}, HERMES
\cite{Airapetian:2006vy}, SLAC E143 \cite{Abe:1998wq}, E155
\cite{Anthony:2000fn} and CLAS \cite{Prok:2014ltt} Collaborations,
at $q^2 > 1$ (GeV/c)$^2$. Also, we consider the very recent data
(2017) from the COMPASS Collaboration \cite{Aghasyan:2017vck}, where
the photon virtuality is $q^2 < 1$ (GeV/c)$^2$, while $4 \times
10^{-5} < x < 4 \times 10^{-2}$. The chi-square value per degree of
freedom that we obtain for our best fit corresponding to the
conformal model is $\chi_{d.o.f.}^2 = 1.140$, while for the
hard-wall model our fit gives $\chi_{d.o.f.}^2 = 1.074$. In both
cases we fit the structure function $g_1$ against data from the
COMPASS Collaboration \cite{Aghasyan:2017vck}.  Thus, our
predictions lead to a very good fit as we shall discuss in detail in
section 5. Also we have calculated the structure function $F_3$.

The holographic dual model corresponding to the planar limit of
${\cal{N}}=4$ SYM theory is represented by a solution of type IIB
supergravity on AdS$_5\times S^5$. The metric can be written as
\begin{equation}
ds^2 =
\frac{R^2}{z^2}\left(\eta_{\m\n}dx^\m dx^\n+dz^2\right)+ R^2
d\Omega_5^2 \, .
\label{metric}
\end{equation}
with radius $R=(4 \pi \, \lambda \, \alpha'^2)^{1/4}$. The
ten-dimensional indices are denoted by $M, N, \dots =0, \dots, 9$,
the AdS$_5$ ones are $m, n,\dots=0,\dots, 4$, the flat
four-dimensional indices are $\m,\n,\dots=0,\dots, 3$, while the
$S^5$ indices are $a,b,\dots=1,\dots,5$. The region $z\rightarrow 0$
corresponds to the UV. In the IR we assume the cut-off
$z_0=1/\Lambda$.

In \cite{Kovensky:2017oqs} we have calculated holographically the
structure function $F_3(x,q^2)$ for glueballs of ${\cal {N}}=4$ SYM
theory. This has also been done at strong coupling and at low $x$.
Other very interesting developments from first principles
calculations for scalar and polarized vector mesons have been done
in \cite{Koile:2011aa,Koile:2013hba,Koile:2014vca,Koile:2015qsa}, as
well as $1/N$ corrections for glueballs \cite{Jorrin:2016rbx},
scalar mesons \cite{Kovensky:2016ryy} and vector mesons
\cite{Kovensky:2018}. The development of a unified description of
the Regge physics and the BFKL Pomeron using the AdS/CFT duality has
been done in \cite{Brower:2006ea}. Further developments including
the eikonal approach, have been presented in
\cite{Brower:2007xg,Brower:2007qh,Cornalba:2006xm,Cornalba:2007zb,
Hatta:2007he,Nishio:2011xz,Costa:2012fw,Watanabe:2012uc,Costa:2013uia,Nally:2017nsp}.
Other aspects of applications of the AdS/CFT correspondence to DIS
processes can be found in
\cite{BallonBayona:2007qr,BallonBayona:2008zi,BallonBayona:2009uy,
BallonBayona:2010ae,Bayona:2011xj,Ballon-Bayona:2017vlm}.

The work is organized as follows. In section 2 we focus on the
calculation of the symmetric structure functions for spin-$1/2$
hadrons. In sections 2 and 3 we calculate the structure functions
both from the heuristic point of view and from the type IIB
superstring theory scattering amplitudes. All this corresponds to
low but not exponentially low $x$. In section 4 we consider the
calculation of $g_1$ in the exponentially small region of the
Bjorken parameter, extending the BPST Pomeron techniques to the
Reggeized gauge field. In section 5 we analyze our results and make
comparison to the existing experimental data for $g_1$.

%
\section{DIS from spin-$1/2$ hadrons at low $x$: the graviton exchange contribution}
%

In this section we focus on the calculation of the symmetric
structure functions for DIS of charged leptons from spin-$1/2$
hadrons at low $x$. The dual holographic calculation involves a
graviton exchange in the $t$-channel as shown in figure 4.a (also
see figure 3).
\begin{figure}[ht]
\centering
\begin{subfigure}[]{
\includegraphics[scale=1]{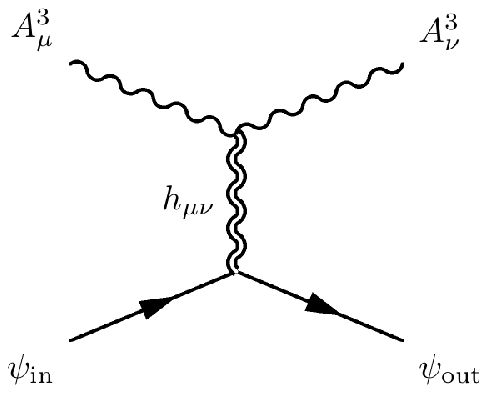}}
\end{subfigure}
\begin{subfigure}[]{
\includegraphics[scale=1]{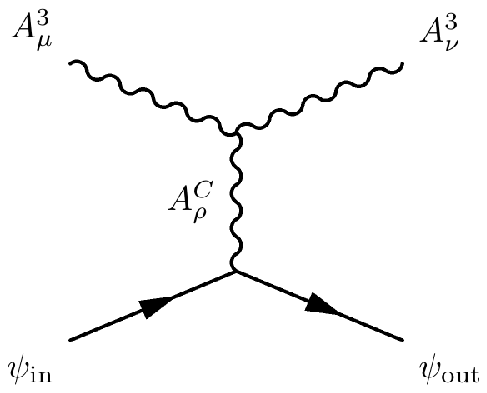}}
\end{subfigure}
\label{DISFCS} \caption{\small $t$-channel holographic dual
representation of forward Compton scattering at tree-level. Figure
(a) shows the exchange of a graviton in the AdS bulk, leading to the
calculation of symmetric structure functions. Figure (b) indicates
the Chern-Simons interaction in top vertex and the propagation of a
bulk-to-bulk gauge field, leading to the anti-symmetric structure
functions.}
\end{figure}

For the Bjorken parameter $x$ within the parametric region
$\lambda^{-1/2} \ll x < 1$, at strong coupling and for large $N$,
double-trace operators dominate the operator product expansion (OPE)
of two electromagnetic currents inside the hadron. The scattering is
produced from the entire hadron. In this regime the holographic dual
description can be done in terms of the calculation of the
$s$-channel in type IIB supergravity schematically shown in figure
2. Beyond that regime, at low $x$ (more precisely when $x \ll
\lambda^{-1/2}$) the holographic dual description of DIS requires
considering the dynamics of type IIB superstring theory on the
AdS$_5\times S^5$ background. In particular, for values of the
Bjorken parameter in the $\exp{\left(-\lambda^{1/2}\right)} \ll x
\ll \lambda^{-1/2}$ range it is possible to carry out the
holographic dual description in terms of scattering amplitudes of
closed strings propagating in ten-dimensional spacetime
\cite{Polchinski:2002jw}. In fact, as argued in
\cite{Polchinski:2002jw, Brower:2006ea}, the dominant $t$-channel
contribution to the DIS process is well described by local
flat-space scattering amplitudes. Therefore, an effective Lagrangian
can be built out from the local string theory scattering amplitude.
Then, in order to obtain the dual FCS amplitude from which one can
derive the structure functions, we have to take the imaginary part
and integrate over the full AdS$_5\times S^5$. Since we focus on a
spin-$1/2$ hadron, the dual closed string modes are associated with
the ten-dimensional dilatino $\Psi(x^M)$.

On the other hand, the relevant effective Lagrangian can also be
constructed in a heuristic way
\cite{Kovensky:2017oqs}\footnote{Strictly speaking this method only
gives the AdS$_5$ contribution, thus we have to multiply by {\it ad
hoc} contribution from the integration on $S^5$ which only gives an
overall factor. The dependence on the $S^5$ radius is accounted for
by using dimensional analysis.} (also see \cite{Gao:2009se}), which
basically involves two steps. Firstly, we have to consider the
five-dimensional supergravity interactions together with the
graviton propagator. Secondly, we need to combine them by taking a
local limit and interpreting the resulting expression of the
propagator as coming from the $\alpha'$-dependent pre-factor of the
string theory scattering amplitude\footnote{Details are given in
reference \cite{Kovensky:2017oqs}.}. This leads to the so-called
ultra-local approximation of the scattering amplitude.

In both frameworks, {\it i.e.} the heuristic and the first-principle
gauge/string theory dual approaches, it is possible to calculate the
symmetric structure functions of the spin-$1/2$ hadron. In the
string theory scattering amplitude approach, the DIS process is
related to the choice of the external modes: while the
ten-dimensional dilatino field is given by a Neveu-Schwarz-Ramond
(NS-R) field, we consider the photon to be a particular polarization
state of the graviton NS-NS mode as in \cite{Polchinski:2002jw}. In
the heuristic approach, the external states are described by
Kaluza-Klein (KK) modes corresponding to ten-dimensional modes. At
low $x$, the four-dimensional center-of-mass (CM) energy $s$ is very
high since
\begin{equation}
s \equiv - (P+q)^2 \approx -q^2 - 2 P\cdot q = - q^2 \left(1 -
\frac{1}{x}\right) \approx \frac{q^2}{x},
\end{equation}
where we have used the fact that in this regime $-P^2 \ll q^2 \ll
-P\cdot q$. This implies that the ten-dimensional Mandelstam
variable $\tilde{s}$ becomes very large. Thus, the leading
contribution to the scattering process comes from the $t$-channel
exchange. When the exchanged field carries spin $j$ this
contribution is proportional to $\tilde{s}^j$. Consequently, the
dominant process in this context is the $t$-channel Reggeized
graviton exchange where $j \approx 2$.

In the next subsection we derive an effective Lagrangian in a
heuristic approach. In subsection 2.2 we carry out the formal
derivation of the Lagrangian starting from the four-point type IIB
superstring theory scattering amplitude, with one NS-R, one R-NS and
two NS-NS fields. In subsection 2.3 we explicitly obtain the
symmetric structure functions.

%
\subsection{Heuristic derivation of the effective Lagrangian}
%

In order to construct the heuristic effective Lagrangian leading to
the symmetric part of the hadronic tensor $W^{\m\n}$ at low $x$ we
need to consider a $t$-channel five-dimensional $SU(4)$ gauged
supergravity tree-level diagram, as shown in figure 4.a. This
maximally supersymmetric supergravity is obtained from dimensional
reduction of type IIB supergravity on $S^5$
\cite{Kim:1985ez,Gunaydin:1984qu,Pernici:1985ju,
Gunaydin:1985cu,Freedman:1998tz}. This spontaneous compactification
of type IIB supergravity leads to a five-dimensional Chern-Simons
term \cite{Gunaydin:1984qu,Pernici:1985ju, Gunaydin:1985cu} that
will be very important in the calculation of antisymmetric structure
functions described in section 3. We will follow closely the steps
described in our previous paper \cite{Kovensky:2017oqs}, however
there is a crucial difference now, namely: instead of using a
dilaton wave-function, in the present heuristic case we must
consider the wave-function of a dilatino field $\psi(x, z)$,
representing the spin-$1/2$ hadron.

The relevant part of the maximally supersymmetric supergravity
action on AdS$_5$, with indices $m,n=0,...,4$, is given by the
expression \cite{Gunaydin:1985cu}
\bea
S_{5d} = \frac{1}{2\kappa^2_5}\int d^5x
\sqrt{-g_{AdS_5}}\left({\cal{R}} - \bar{\psi} \, \gamma^m
D_m\psi-\frac{1}{4}\left(F_{mn}^A\right)^2+\cdots \right) \, ,
\label{s5d}
\eea
where $2\kappa_5^2 = 16 \pi^2/N^2$ is the Newton constant in five
dimensions (we set $R=1$), $F^A_{mn}$ is the non-Abelian gauge field
strength associated with the gauge field $A_m^A$, and ${\cal{R}}$ is
the Ricci scalar in five dimensions. Also we use the definition
$\gamma^m = e_{\hat{m}}^m \, \gamma^{\hat{m}}$, where
$\gamma^{\hat{m}}$ are the flat-space Dirac matrices
$({\hat{m}}=0,...,4)$ and $e_{\hat{m}}^m$ is the vielbein. Dots
include kinetic and interaction terms which are not relevant for our
present analysis.

At high energy the leading diagram is given by the $t$-channel
exchange of a graviton. Since the graviton couples to the
energy-momentum tensors $T_{mn}^\psi$ and $T_{mn}^A$ given
by\footnote{Note that the fluctuations of the fields are normalized
with an extra factor $\sqrt{2}\kappa_5$. We only write the quadratic
terms of the energy-momentum tensors.}
\bea
T_{mn}^\psi =  \bar{\psi} \, \gamma_{(m}\der_{n)}\psi \;, \qquad
T^A_{kl}= g^{pq}F_{kp}F_{lq} - \frac{1}{4}g_{kl}F_{pq}F^{pq} \, ,
\eea
the corresponding amplitude has the form
\begin{equation}
{\cal{A}}= \kappa_5^2 \int d^5x \, d^5x' \, T_{mn}^\psi(x) \,
G^{mnkl}(x,x') \, T_{kl}^A(x') \, ,
\end{equation}
where $G^{mnkl}(x,x')$ denotes the AdS$_5$ graviton propagator whose
relevant terms can be expressed as
\begin{equation}
G^{mnlk}(x,x')=\left(g^{mk}g^{nl}+g^{ml}g^{nk} - \frac{2}{3}
g^{mn}g^{kl}\right) G_{\mathrm{grav}}(x,x') + \dots \, ,
\end{equation}
being $G_{\mathrm{grav}}(x,x')$ some function whose explicit form we
dot not need.

Gathering all the information we obtain the following integrand
\beq
T_{mn}^\psi(x) \, G^{mnlk}(x,x') \, T^A_{kl}(x) =  2 \,
G_{\mathrm{grav}}(x,x') \, F_{mp}(x') \, F_{n}^p(x') \,
\bar{\psi}(x)\gamma^{n}\der^{m}\psi(x) \, , \label{LLOsugra}
\eeq
plus ${\cal{O}}(t)$ terms. We only consider the leading terms in
$\tilde{s}=-\tilde{u}$, since in the $x \ll \lambda^{-1/2}$ regime
we have $\tilde{s} \gg \tilde{t}$.

In order to obtain the effective action one would have to integrate
the effective Lagrangian obtained from equation (\ref{LLOsugra})
over the full AdS$_5\times S^5$. The sphere reduction gives a
numerical constant $C$. Then, we need to multiply it by the
superstring theory pre-factor
\begin{equation}
\tilde{s}^2 \, {\cal{G}}(\al',\tilde{s},\tilde{t},\tilde{u}) = -
\frac{\al'^3 \tilde{s}^2}{64} \prod_{\chi =
\tilde{s},\tilde{t},\tilde{u}} \frac{\Gamma \left(-\al' \chi
/4\right)}{\Gamma\left(1+\al' \chi /4\right)} \, .
\end{equation}
The effective action is
\bea
S_{{\mathrm{eff}}}^{{\mathrm{(S)}}} = 2 \,\kappa_5^2\,
\textrm{Im}\left[\tilde{s}^2 \,
{\cal{G}}(\al',\tilde{s},\tilde{t},\tilde{u})\right] C\int
d^5x\sqrt{g_{AdS_5}}\ F_{mp}F_n^{\phantom{n}p}\,
\bar{\psi}\gamma^{(m}\der^{n)}\psi \, . \label{SheurSim}
\eea
By plugging the solutions for $\psi(x^\mu, z)$ and $A_m(x^\mu, z)$
in equation (\ref{SheurSim}) we can evaluate the on-shell action and
then take its imaginary part. This leads to the dilatino (symmetric)
structure functions that will be calculated in subsection 2.3. In
the next subsection we show how to derive the effective action from
first principles, starting from the scattering amplitude of four
closed strings in type IIB superstring theory.

%
\subsection{Derivation from the string theory scattering amplitude}
%

In the $e^{-\sqrt{\lambda}}\ll x\ll \lambda^{-1/2}$ regime we can
obtain the spin-$1/2$ hadronic tensor by calculating a certain
tree-level four-point string theory scattering amplitude in
ten-dimensional flat-space. This was motivated in
\cite{Polchinski:2002jw}. Once the local flat-space amplitude is
obtained one can derive an effective Lagrangian, which is then
integrated over the AdS$_5\times S^5$ space after the inclusion of
the curved-space wave-functions of the dilatinos and the
gravi-photons. The five-dimensional spin-$1/2$ and gauge fields
which we have used in the previous section are specific KK modes of
these ten-dimensional excitations reduced on $S^5$. The external
states are given by two dilatinos and two gravitons. The details of
the decomposition are given below. In other words, we are interested
in a closed string amplitude with two modes from the NS-R sector and
the other two from the NS-NS sector.

Following the KLT relations \cite{Kawai:1985xq,Becker:2015eia} the
closed-string theory scattering amplitude factorizes in terms of
open-string amplitudes as
\begin{equation}
{\cal{A}}(1,2,\tilde{3},\tilde{4}) = 4\ i\ \kappa_{10}^2 \,
{\cal{G}}(\alpha',\tilde{s},\tilde{t},\tilde{u})\,
K_{op}^{\mathrm{bos}}(1,2,3,4) \otimes
K_{op}^{\mathrm{fer}}(\tilde{3},1,2,\tilde{4}) \, ,
\end{equation}
where $K_{op}$ are open string kinematic factors. Particle numbers
with a tilde indicate fermionic modes. For these particular
combinations of modes these factors can be found in
\cite{Becker:2015eia,Schwarz:1982jn}. The relevant terms take the
form
\begin{equation}
K_{op}^{\mathrm{bos}}(1,2,3,4) =
\xi_1^{M}\xi_2^{N}\xi_3^{P}\xi_4^{Q} \left[-1/4\,
\tilde{s}\,\tilde{u}\, \eta_{MN}\eta_{PQ} + \cdots \right] \; ,
\label{kopbos}
\end{equation}
and
\bea
K_{op}^{\mathrm{fer}}(\tilde{3},1,2,\tilde{4}) =
\xi_1^{M'}\xi_2^{N'}\bar{u}_3^\alpha u_4^\beta
\left[\tilde{s}\left(k^2_{M'}(\Gamma_{N'})_{\alpha\beta} -
k^1_{N'}(\Gamma_{M'})_{\alpha\beta} -
\eta_{M'N'}(\Gamma^{P})_{\alpha\beta} k^2_{P} \right) + \cdots
\right] \, , \label{kopfer}
\eea
where dots indicate sub-leading terms in the dual DIS process.
$\xi_i$ and $u_i$ are the boson and fermion polarizations,
respectively, while $\Gamma^N$ indicates the ten-dimensional gamma
matrices. The spinor indices are denoted by $\alpha, \beta$ and the
ten-dimensional bosonic indices are denoted by $M, N$. In the
notation of equations (\ref{kopbos}) and (\ref{kopfer}) the
ten-dimensional Mandelstam variables are defined as
\begin{equation}
\tilde{s}=-(k_1 + k_4)^2 \ , \ \tilde{t}=-(k_1 + k_2)^2  \ , \
\tilde{u}=-(k_1 + k_3)^2 \, ,
\end{equation}
where $k_1$ and $k_2$ are the momenta associated to the bosonic
modes, while $k_3$ and $k_4$ are the ones associated to the
fermionic modes. Also, the closed-string graviton and dilatino
polarizations are given by \cite{Garousi:1996ad}
\beq
h^{MN}_i \equiv \xi_i^M \otimes \xi_i^N \;,\quad
(\Gamma^M)_\beta^\alpha \Psi^\beta_i \equiv u_i^\alpha \otimes
\xi_i^M\, . \label{closedpolarization}
\eeq
Thus, to leading order in $\tilde{s}$ the corresponding amplitude
becomes
\bea
{\cal{A}}(1,2,\tilde{3},\tilde{4})= 4 \, i \, \kappa_{10}^2 \,
{\cal{G}} \, \tilde{s}^2 \, \overline{\Psi}_3\left[ \tilde{u}(h_1\cdot
h_2)\slashed{k_1} + \tilde{s}(h_1\cdot h_2)\slashed{k_2}\right.
\label{A1}\\
\left.+ 2\ \tilde{u}(k_2\cdot h_1\cdot h_2\cdot\Gamma)+ 2\
\tilde{s}(k_1\cdot h_2\cdot h_1 \cdot\Gamma)\right]\Psi_4 \,.\nn
\eea
In this expression we can set $k^1_M h_2^{MN}=k^2_M h_1^{MN}=0$
since both graviton states correspond to the ingoing dual photon and
its complex conjugate associated with the outgoing one,
respectively. Thus, from now on we will neglect the first two terms
in equation (\ref{A1}). Then, the effective Lagrangian associated
with this scattering amplitude can be written by replacing momenta
with derivatives, giving the following structure
\begin{equation}
 - i \, \kappa^2 \, (\partial_P
h_{MN}) \, \, (\partial_Q h^{MN}) \, \, \overline{\Psi} \,
\Gamma^{(P} \, \partial^{Q)}\Psi \, .
\label{L10dsym}
\end{equation}

Next, we need to obtain a curved-space version of (\ref{L10dsym})
and rewrite it in terms of the five-dimensional fields. The
decomposition of the fields is given by \cite{Kim:1985ez}
\begin{equation}
\Psi (x^m,\Omega) = \sum_{\Delta} \psi_\Delta (x^m) \otimes
\eta_\Delta (\Omega) \, , \,\,\,\, h^{m a} = \sum_k A^m_{k}(x^n) \,
Y^a_{k}(\Omega) \, , \label{expansion}
\end{equation}
where $\Delta$ and $k$ are integers, $\eta_\Delta (\Omega)$ are
eigenfunctions of the Dirac operator on $S^5$ and $Y^a_{k}(\Omega)$
are the corresponding vector spherical harmonics. In the holographic
dual DIS process calculation, we focus on a particular value of
$\Delta$ (note that henceforth we write $\psi_\Delta \equiv \psi$).
Also, the massless gauge field has the lowest vector spherical
harmonics, which are given by the Killing vectors $K^A_a$ on $S^5$.
Moreover, when considering gauged supergravity both vector fields
carry a gauge group index\footnote{If the full isometry group of the
sphere $SO(6)\sim SU(4)$ is gauged the index $A$ runs from $1$ to
$15$.} $A$. Thus, we can write
\begin{equation}
\Psi(x^M) \rightarrow \psi(x^m) \otimes \eta (\Omega) \, ,
\,\,\,\,\, h_{MN} \rightarrow h_{m a} = A_{(m}^A(x^n) \, K^A_{a)} \,
. \label{Psiaaa}
\end{equation}
Plugging these expressions in the effective Lagrangian and
integrating over AdS$_5\times S^5$ we obtain the effective on-shell
action (\ref{SheurSim}), where $C$ is defined by considering the
normalization condition
\bea
\int d^5\Omega \, \sqrt{g_{S^5}} \, \bar{\eta}(\Omega) \,
\eta(\Omega) \, K^a \, K_a = C  \, . \label{NormArm}
\eea
%

%
\subsection{Symmetric structure functions}
%

In this section we calculate the symmetric structure functions of a
spin-$1/2$ hadron (which is assumed to be dual to a dilatino bound
state as in \cite{Polchinski:2002jw}) in the $e^{-\sqrt{\lambda}}\ll
x \ll \lambda^{-1/2}$ regime. We follow the conventions of reference
\cite{Polchinski:2002jw}. We consider the AdS$_5$ metric given in
\eqref{metric}. In the hard-wall model a radial cut-off is included
at $z_0=\Lambda^{-1}$, in order to account for the IR confinement
scale $\Lambda$ in the dual field theory. For energy larger than
$\Lambda$ the theory becomes approximately conformal.

In order to compute the hadronic tensor we need to obtain the
effective action (\ref{SheurSim}) evaluated on-shell. Since the AdS
process is dual to the FCS, the imaginary part gives the DIS
hadronic tensor as follows
\begin{equation}
S_{{\mathrm{eff}}}^{{\mathrm{(S)}}} \equiv n_\mu n^*_\nu
\,{\mathrm{Im}} \, \left[T^{\mu\nu}_{(S)}\right] = \frac{1}{2\pi}
n_\mu n^*_\nu \, W^{\mu\nu}_{(S)} \, .
\end{equation}
The incoming and outgoing gauge fields are given by the
non-normalizable solutions of the Einstein-Maxwell equations in AdS.
By imposing the appropriate boundary conditions
\begin{equation}
A_\m^3 (z\rightarrow 0) = n_\mu e^{i q \cdot x} \, , \,\,\,\,\
A_z^3 (z\rightarrow 0) = 0 \, ,
\end{equation}
the solutions are given by
\begin{equation}
A_\m^3 = n_\m \, e^{i q \cdot x} \, q z \, K_1\left(qz\right) \, , \,\,\,\,\,
A_z^3 =  i (n\cdot q) \, e^{i q \cdot x} \, z \, K_0(qz) \, ,
\label{Amu}
\end{equation}
where $K_i$ denotes the Bessel functions of the second kind. Note
that without loss of generality we can choose a transversal
polarization for the virtual photon. Thus, from now on we take $n
\cdot q =0$ and in particular we set $A_z^3=0$.

Now, let us consider the dilatino. We briefly describe the
corresponding type IIB supergravity solution following
\cite{Polchinski:2002jw}. In the conformal region we can write the
dilatino wave-function as in equation (\ref{Psiaaa}). The $\psi(x,
z)$ solution satisfies the Dirac equation in five dimensions.
Factorizing out the spinor harmonic $\eta (\Omega)$ on the sphere,
the five-dimensional solution with four-momentum $P_{\mu}$ is
\beq
\psi = e^{iP\cdot x} \, C' \,
z^{5/2}\left[J_{\tau-2}(Pz)P_{+}+J_{\tau-1}(Pz)P_{-}\right] u(P) \,
, \label{PsiaaaC}
\eeq
where $C'$ is a normalization constant, $\tau=\Delta-1/2=mR+3/2$ is
the twist of the corresponding QFT operator, and the
four-dimensional chirality projectors are defined as
$P_{\pm}\equiv\frac{1}{2}\left(1\pm\gamma^5\right)$ with $\gamma^5
\equiv \gamma^{\hat{z}}$. Also, $u$ and $\overline{u}$ are Dirac
spinors in four dimensions.

The leading terms in the near-boundary expansion are given
by\footnote{The second terms in both equations (\ref{psi}) give
$P^2/q^2$ sub-leading contributions to the symmetric structure
functions, thus we will not consider them in the following
calculations.}
\bea
\psi_i &\approx& e^{iP\cdot x}\frac{c_i}{\Lambda^{3/2}}
(z/z_0)^{\tau+1/2}\left[ P_+ + \frac{Pz}{2(\tau-1)}P_{-} \right] u_i(P) \, , \nn \\
\overline{\psi}_i &\approx& e^{-iP\cdot x}\frac{c_i^{*}}{\Lambda^{3/2}}
(z/z_0)^{\tau+1/2}\,\overline{u}_i(P)\left[ P_- + \frac{Pz}{2(\tau-1)}P_{+} \right] \, ,
\label{psi}
\eea
where $c_i$ is some dimensionless constant.

Since we consider the $\tilde{t} \to 0$ and  $\tilde{s} \to \infty$
limit, we can expand the string theory scattering amplitude
pre-factor as in \cite{Polchinski:2002jw}. Thus, by taking the
imaginary part we can rewrite it as a sum over the excited states in
the form
\begin{equation}
{\mathrm{Im}}_{{\mathrm{exc}}}\left[G(\al',\tilde{s},\tilde{t},\tilde{u})\
\tilde{s}^2\right]|_{\tilde{t}\rightarrow 0} = \frac{\pi
\al'}{4}\sum_{m=1}^{\infty} \delta \left(m - \frac{\al'
\tilde{s}}{4}\right) (m)^{\al' \tilde{t}/2} \, , \label{IMG}
\end{equation}
where the last factor can be ignored in the region of interest since
$\al' \tilde{t} \sim {\cal {O}}(\lambda^{-1/2})$. This sum can be
approximated by an integral for $x \ll \lambda^{-1/2}$. Recall that
the relation between the ten-dimensional Mandelstam variables and
the four-dimensional ones is
\begin{equation}
\al' \tilde{s} \approx \al' s \, z^2/R^2 \, , \label{sexp}
\end{equation}
plus corrections from the radial and $S^5$ coordinates which can be
neglected.

Note that once the fermion solutions are inserted, the objects with
spinor indices in the leading term give a factor
\begin{equation}
\overline{u}_i P_- \gamma_{\hat{\mu}} P_+ u_i = \overline{u}_i
\gamma_{\hat{\mu}} P_+ u_i = -i (P_\mu + S_\mu) \, ,
\end{equation}
where $S_\mu$ is the spin polarization vector. However, the second
term is actually misleading and should be omitted. The graviton
exchange of the dual calculation corresponds to the energy-momentum
tensor term in the current-current OPE (on the QFT side). Thus,
terms proportional to $S_\mu$ should not be present in the
expectation value. From the holographic dual approach it is
necessary to go back to the full expression for the spin-$1/2$
solution and "undo" the local approximation for the $t$-channel
diagram. Then, the $z$-integral (with the correct integration limits
$0 <z <z_0$) can be computed in the relevant low-momentum limit of
the graviton mode, giving a vanishing result (as opposed to the
contributions proportional to $P_\mu P_\nu$). The details of the
computation are very similar to those of reference \cite{Gao:2009ze}
where the authors study the elastic form factors for the conserved
current. Note that this observation implies that in the string
theory regime the structure functions $g_{3,4,5}(x,q^2)$ will vanish
at leading order in the $1/N$ expansion.

Now, plugging all these elements in the effective action and
carrying out the integrals over AdS$_5\times S^5$ we find
\bea
n_\mu n^{*}_\nu T^{\mu \nu}_{\mathrm{(S)}} &=& n_\m^* n_\n
\frac{\pi\, |c_i|^2 C}{2\sqrt{4\pi\lambda}}
\left(\frac{\Lambda^2}{q^2}\right)^{\tau-1} q^{-2} \times
\left[\eta^{\mu\nu}\frac{(P\cdot q)^2}{q^2} I_{1,2\tau+3} + P^\mu
P^\nu (I_{0,2\tau+3} + I_{1,2\tau+3})\right] \label{ImSsinx} \nn \\
&&
\eea
where
\beq
I_{j,n} = \int_0^\infty dw\ w^n K_j^2(w) =
2^{n-2}\frac{\Gamma(\n+j)\Gamma(\n-j)\Gamma(\n)^2}{\Gamma(2\n)} \ ,
\ \n=\frac{1}{2}(n+1) \ , \ I_{1,n}=\frac{n+1}{n-1}I_{0,n} \, .
\eeq
Writing the above expressions in terms of the Bjorken parameter
$x=-\frac{q^2}{2(P\cdot q)}$ and comparing with the structure of
hadronic tensor (\ref{w}), we obtain the following symmetric
structure functions for the spin-$1/2$ hadron
\beq
F_1\left(x,q^2\right) =
\frac{1}{x^2}\left(\frac{\Lambda^{2}}{q^2}\right)^{\tau-1}\frac{\pi^2|c_i'|^2
C}{4(4\pi\lambda)^{1/2}}I_{1,2\tau+3} \ \ , \ \
F_2\left(x,q^2\right) =  2x\frac{2\tau+3}{\tau+2}F_1(x,q^2) \, ,
\label{F222}
%
\eeq
together with $g_3 = g_4 = g_5 = 0$. Note that the $x$ and $q^2$
dependence of $F_1$ and $F_2$ structure functions agree with the
ones obtained by Polchinski and Strassler for the (scalar) glueball
in the same parametric regime, by interchanging $\Delta$ and $\tau$.
Also, the above equation (\ref{F222}) gives a generalization of the
Callan-Gross relation of a spin-$1/2$ hadron. In addition, there are
no contributions to the antisymmetric structure functions coming
from the $t$-channel graviton exchange. In the next section we shall
see how they appear in a different way.

%
\section{DIS from spin-$1/2$ hadrons at low $x$: the gauge field exchange contribution}
%

We now describe the calculation of the antisymmetric contributions
to the hadronic tensor and derive the corresponding structure
functions for polarized DIS of charged leptons from spin-$1/2$
hadrons at low $x$ and at strong 't Hooft coupling in the large $N$
limit of the ${\cal {N}}=4$ $SU(N)$ SYM theory with an IR cut-off.
The corresponding dual holographic calculation is dominated by the
exchange of a gauge field in $t$-channel within the AdS space, as
shown in figure 4.b. A heuristic analysis has been done in
\cite{Hatta:2009ra} for the $F_3$ function, while in our previous
paper \cite{Kovensky:2017oqs} we have done a first principles
calculation from type IIB superstring theory four-point scattering
amplitude for glueballs.

From a heuristic viewpoint, one can understand that the
antisymmetric contribution arises due to the Chern-Simons term in
the five-dimensional $SU(4)$ gauged supergravity action. In our
conventions, it can be written as
\beq
S_{CS} = \frac{i\ \kappa}{96\pi^2}d_{ABC}\int d^5x \, \vep^{mnopq}
\, A^A_m \, \der_n A^B_o \, \der_p A^C_q \, , \label{scs}
\eeq
where $A,B,C$ stand for the $SU(4)$ gauge group indices,
$\vep^{mnopq}$ is the Levi-Civita symbol, $k$ an integer and
$d_{ABC}$ is the completely symmetric symbol. By coupling the matter
fields to the $A_m^3$ gauge field (the gravi-photon), throughout the
exchange of a spin-one field, the antisymmetric $F_3$ and $g_1$
structure functions are obtained. The exchanged gauge field cannot
be $A_m^3$ because the interaction term includes the $d_{ABC}$
symmetric symbol \cite{Baguet:2015sma,Kovensky:2017oqs}.

In fact, there are two tree-level $t$-channel Feynman diagrams
contributing to the coupling of the dilatino to the Chern-Simons
term. One involves the exchange of a gauge field $A^C_m$ associated
to an $S^5$ isometry. This coupling also appears in the dilaton DIS
\cite{Kovensky:2017oqs}. The second diagram comes from the so-called
Pauli term, which was discussed in the holographic dual description
of DIS in reference \cite{Gao:2010qk}, but only for the
$\lambda^{-1/2} \ll x < 1$ regime. In the non-Abelian case, it takes
the form
\beq
S_{P} = \beta^A \int d^5x \, \sqrt{-g_{AdS}} \, F_{mn}^A \,
\bar{\psi} \left[\gamma^m,\gamma^n\right] \psi \, .
    \label{Lpau}
\eeq
for some constants $\beta^A$. The interaction also occurs through
the exchange of a gauge field. However, note that it is present even
when the dilatino is not charged under the usual isometries.

In terms of the superstring scattering amplitudes, the antisymmetric
contribution leading to $F_3$ and $g_1$ comes from the R-R sector of
the closed string. This occurs in a way similar to the dilaton case.
Recall that the massless five-dimensional gauge field $A_m$ that
emerges after the $S^5$ spontaneous compactification is a linear
combination of the graviton $h_{MN}$ and a particular mode of the
R-R self-dual five-form field strength \cite{Kim:1985ez}. Therefore,
it is important to consider that the incoming dual non-Abelian gauge
states contain modes from both the NS-NS and the R-R sectors.

%
\subsection{Heuristic derivation of the effective Lagrangian}
%

We are now interested in the calculation of the $t$-channel
supergravity process at tree-level, but in this case the exchanged
field has spin one. By looking at the figure 4.b the top-vertex
interaction is given by non-Abelian Chern-Simons term (\ref{scs}).
The two diagrams that we will analyze are schematically represented
by the Feynman diagram shown in figure 4.b.

The Chern-Simons term involves the full set of non-Abelian gauge
fields $A^C_m$ associated to the $SU(4)$ symmetry group. We focus on
processes where the non-normalizable mode dual to the virtual photon
is $A^3_m$, since this mode couples to the electromagnetic current
of the AdS boundary gauge theory. The completely symmetric symbol
$d_{ABC}$ in equation (\ref{scs}) is then restricted to the form
$d_{33C}$. Thus, it is easy to see that the $t$-channel propagating
gauge boson $A^C_m$ can only have color numbers $C=8$ or $C=15$ (see
for example \cite{Hatta:2009ra, Kovensky:2017oqs}). These indices
are associated with two diagonal matrices in the Lie algebra of
$SU(4)$. The idea is to write a heuristic effective Lagrangian, for
which we have to consider the corresponding $t$-channel gauge field
propagator and couple the $A^C_m$ coming from the Chern-Simons
current to the dilatino current. The amplitude can be written as
\begin{equation}
{\cal{A}} = \kappa_5^2 \, \int d^5x \, d^5x'\ {\cal{J}}^{m}_C(x) \,
G_{mn}^{CD}(x,x') \, J^{n}_{D}(x') \, , \label{acs}
\end{equation}
where $ {\cal{J}}^{m}_C$ denotes the Chern-Simons current and
$J^{n}_{D}$ is given by
\beq
{\cal{J}}^{m}_{C}(x) = \frac{i}{6} \, d_{ABC} \, \vep^{mnopq} \,
\der_n A_o^A \, \der_p A_q^B \, , \quad  J^{n}_{D} (x') = -
{\cal{Q}}_D \, \bar{\psi}\gamma^n\psi \, , \label{current}
\eeq
respectively, while $G_{mn}^{CD}(x,x')$ is the gauge field
propagator in AdS$_5$, whose relevant part at high energy can be
expressed as
\beq
G_{mn}^{CD}(x,x')=g_{mn} \, \delta^{CD} \, G_{\mathrm{gauge}}(x,x')
+ \cdots \, , \label{prop}
\eeq
with $G_{\mathrm{gauge}}(x,x')$ being some function which is not
relevant for the present calculation. The charge $Q_D$ in the
dilatino current is related to the eigenvalue equation for the
ten-dimensional wave-function $\Psi(x^m,\Omega)$
\begin{equation}
K^{a}_D \der_a \Psi(x^m,\Omega) = - {\cal{Q}}_D \Psi(x^m,\Omega) \,
,
\end{equation}
being $D=8,15$ the Lie algebra indices corresponding to the matrices
$T_D$.

The rest of the computation is analogous to the symmetric case.
After including the string pre-factor times a constant $\tilde{C}$
coming from $S^5$ integration, and performing the curved-space
integrals, we obtain
\beq
S_{\mathrm{eff}}^{\mathrm{(A)}} = -i \, \frac{1}{6}{\cal{Q}}^C
d_{ABC}\ {\mathrm{Im}} \left[{\cal{G}} \, \tilde{s}^2\right]
\tilde{C} \, \int d^5x \, \vep^{mnopq} \, \der_m A^{A}_n \,
\der_{o}A^{*B}_{p}\ \bar{\psi}\gamma_{q}\psi \, . \label{SasymCS}
\eeq
A similar method can be used for the Pauli term contribution.

%
\subsection{Derivation from the string theory scattering amplitude}
%

Now, we formally derive the effective Lagrangian which permits to
obtain the antisymmetric structure functions from type IIB
superstring theory. For that we first obtain the string theory
scattering amplitude that we need in order to construct the
associated effective Lagrangian relevant for the antisymmetric
contribution. The only difference is that in the present case the
four-point scattering amplitude must contain external states coming
from the R-R sector. The reason for the presence of the R-R sector
is that the massless gauge fields $A_m^C$ of the five-dimensional
$SU(4)$ gauge supergravity are constructed as linear combinations of
two low-lying KK modes on $S^5$, coming from both NS-NS and R-R
string states. The former is a graviton perturbation $h_{MN}$, while
the second one corresponds to a R-R four-form field perturbation
${\cal{C}}_{M_1\cdots M_4}$. This is described in detail in
\cite{Kim:1985ez} and reviewed in our previous work where we have
investigated the dilaton case related to the DIS from glueballs
\cite{Kovensky:2017oqs}.

The relevant four-point amplitudes can be written as one of the two
forms
\begin{equation}
{\cal{A}} \left(\text{R-R} , \text{R-R}, \text{NS-R},
\text{NS-R}\right) \ \ \text{or} \ \ {\cal{A}} \left(\text{NS-NS} ,
\text{R-R}, \text{NS-R}, \text{R-NS}\right), \nonumber
\end{equation}
where the first two external states correspond to the gauge fields
in both cases. The two amplitudes above are important. We explicitly
calculate the first one and show that it leads exactly the effective
action (\ref{SasymCS}) associated with the coupling between the
Chern-Simons term and the minimal coupling of the dilatinos with the
gauge field. Then, we argue why the second amplitude should lead to
the case where this minimal coupling is replaced by the Pauli term.

Next, we want to obtain the scattering amplitude for two NS-R and
two R-R states following the same steps as in section 2.2. Due to
the KLT relations between open and closed superstring amplitudes,
one can see that the amplitude we are interested in is given by
\cite{Kawai:1985xq, Becker:2015eia}
\begin{equation}
{\cal{A}}(\tilde{1},\tilde{2},\textit{3},\textit{4}) = - i\ \kappa^2
{\cal{G}}(\alpha',\tilde{s},\tilde{t},\tilde{u})\,
K_{op}^{\mathrm{fer}}(\tilde{1},\tilde{2},\tilde{3},\tilde{4})\otimes
K_{op}^{\mathrm{fer}}(\tilde{3},1,2,\tilde{4}) \,,
\end{equation}
where the italic numbers stand for the R-R fields. The first
kinematic factor is
\beq
K_{op}^{\mathrm{fer}}(\tilde{1},\tilde{2},\tilde{3},\tilde{4}) =
\frac{\tilde{s}}{2} \,\bar{u}_1\Gamma^{M}u_2\,\bar{u}_3\Gamma_{M}u_4
\, ,
\eeq
and second one is given in equation (\ref{kopfer}). The dilatino
polarizations are given in equation (\ref{closedpolarization}), and
the polarizations of the closed-string four-form field are given in
terms of the open-string ones by
\beq
u_i^\alpha \otimes \bar{u}_i^\beta = ({\cal{C}}_Q
\Gamma_{i(5)})^{\alpha\beta}\,,\quad \mathrm{with}\quad
\Gamma_{i(5)}=({\cal{F}}_i)_{M_1\cdots M_5}\Gamma^{M_1\cdots M_5} \,
,
\eeq
in the conventions of \cite{Becker:2015eia}, being ${\cal{C}}_Q$ the
charge conjugation matrix. After some algebra, we obtain the leading
amplitude in this regime
\bea
{\cal{A}}(\tilde{1},\tilde{2},\tilde{3},\tilde{4}) &=& - i\ \kappa^2
{\cal{G}}(\alpha',\tilde{s},\tilde{t},\tilde{u})\, \tilde{s}^2
\frac{16}{15} ({\cal{F}}_{3})_{M M2\cdots
M_5}({\cal{F}}_4)_{N}^{\phantom{a}M_1\cdots M_5} \bar{\Psi}_1
\gamma^{(N}k_2^{M)}\Psi_2 \, ,
\eea
from which the effective Lagrangian can be constructed. For that
purpose we consider the curved metric together with the
$S^5$-reduction of the different fields, which are rewritten as an
expansion in modes over the $S^5$. The dilatino expansion is given
in equation (\ref{expansion}), while the 5-form field strength
perturbation is \cite{Kim:1985ez, Kovensky:2017oqs, Baguet:2015sma}
\beq
{\cal{F}}_{mnabc} \sim 5 (1+*) \, \partial_{[n} A^{B}_{m]} \,
Z_{abc}^B \, ,\quad {\mathrm{with}}\quad Z_{abc}^A \equiv
\ep_{abcde} \, \nabla^d K^{eA}  \, , \label{pert}
\eeq
where $\ep_{abcde}$ is the Levi-Civita tensor on the sphere.

The charges ${\cal{Q}}^C$ come from the harmonic spinors
transformation under the corresponding $S^5$ isometries, while the
$d_{ABC}$ symbol emerges from a Killing vector identity
\cite{Kovensky:2017oqs}. Then, multiplying by the string pre-factor
we obtain the effective action with the following structure
\beq
-i \, d_{ABC} \, {\cal{Q}}^C\, {\mathrm{Im}} \left[{\cal{G}} \,
\tilde{s}^2\right] \int d^5\Omega \, \sqrt{g_{S^5}} \,
\bar{\eta}(\Omega) \, \eta(\Omega) \, \, \int d^5x \,
\epsilon^{mnopq} \, \der_m A^{A}_n \, \der_o A_p^{B*} \, \bar{\psi}
\gamma_q \psi \,.
\eeq
The dependence on the fields and the Mandelstam variables of this
result fully agrees with equation (\ref{SasymCS}).

The fact that this particular amplitude leads to the effective
action (\ref{SasymCS}) could have been anticipated by looking at the
three-point string theory scattering amplitudes. As it has been
carefully analyzed in our previous paper \cite{Kovensky:2017oqs},
one can see that the ${\cal{A}}({\text{R-R, R-R, NS-NS}}) \sim
{\cal{A}} ({\cal{F}},{\cal{F}},h)$ leads to an interaction term of
the form of the Chern-Simons term in the supergravity action. One
obtains this precisely when the external states have the particular
polarizations indicated above. Thus, the Feynman diagram associated
with the minimal coupling comes from an amplitude where the graviton
state (with one index on AdS$_5$ and the other on $S^5$) propagates
in the $t$-channel. In this heuristic approach, the two incoming
dual gauge fields are modes of the self-dual five-form field
strength. Then, the dilatino and dilaton IR vertices come from the
string theory scattering amplitudes
${\cal{A}}({\text{NS-NS,NS-NS,NS-NS}})$ $\sim {\cal{A}}
(h,\phi,\phi)$ and ${\cal{A}}({\text{NS-NS,NS-R,NS-R}}) \sim
{\cal{A}} (h,\Psi,\Psi)$, respectively.

For the Pauli term, on the other hand, one can use a similar
reasoning. It is not difficult to see that from the three-point
scattering amplitude ${\cal{A}}({\text{R-R,R-NS,NS-R}}) \sim
{\cal{A}} ({\cal{F}},\overline{\psi},\psi)$, supplemented with the
corresponding polarizations, one can derive a five-di\-men\-sional
effective Lagrangian of the form of the Pauli interaction term
(\ref{Lpau}), at least in the AdS$_5$ space. This is so because in
terms of the ten-dimensional fields the effective Lagrangian is of
the form
\begin{equation}
L_{{\cal{F}}\overline{\Psi}\Psi} \propto {\cal{F}}_{MNOPQ} \,
\overline{\Psi} \Gamma^{MNOPQ} \Psi \, ,
\end{equation}
and then we only have to take two indices on AdS$_5$ and the other
three on $S^5$. Therefore, we can consider that in this case the
exchanged gauge field is a five-form field strength mode. Then,
since in this case the top vertex is derived from the Chern-Simons
term, we conclude that it should be possible to explicitly obtain
the effective action coming from the Pauli interaction term by
studying the ${\cal{A}}({\text{NS-NS, R-R, NS-R, R-NS}}) \sim
{\cal{A}}({\cal{F}}, h, \overline{\Psi}, \Psi)$ four-point
amplitude.

%
\subsection{Antisymmetric structure functions}
%
%
%
{\bf Contribution of the Chern-Simons term} \\
%

In this subsection we explicitly derive the antisymmetric structure
functions of the spin-$1/2$ hadron. We need to evaluate the
effective action on-shell, and then use the holographic relation
\begin{equation}
-iS_{\mathrm{eff}}^{\mathrm{(A)}} \equiv n_\mu n^*_\nu
\,{\mathrm{Im}} \left[T^{\mu\nu}_{\mathrm{(A)}}\right] =
\frac{1}{2\pi} n_\mu n^*_\nu W^{\mu\nu}_{\mathrm{(A)}} \, .
\end{equation}
Both the heuristic and the string-amplitude approaches give the same
effective action. Let us consider equation (\ref{SasymCS}). The
AdS$_5$ solutions are given in equation (\ref{Amu}) for the incoming
gauge field $A_m^3$ and in equation (\ref{psi}) for the dilatino.
Using equations (\ref{IMG}) and (\ref{sexp}) to evaluate the
imaginary part of the string pre-factor, we obtain
\beq
n_\mu n^*_\nu \,{\mathrm{Im}}
\left[T^{\mu\nu}_{\mathrm{(A)}}\right] = \varepsilon^{\mu \nu \rho
\sigma} n_\m n_\n^* q_{\rho}\,P_\sigma q^{-2} {\cal{Q}}\ \frac{\pi\,
|c_i|^2}{12\sqrt{4\pi\lambda}}
\left(\frac{\Lambda^2}{q^2}\right)^{\tau-1}  {\cal{I}}_{\tau} \, ,
\label{ImSA}
\eeq
where the charge is defined as ${\cal{Q}} \equiv
d_{33C}{\cal{Q}}^C$. We also define
\begin{equation}
{\cal{I}}_\tau \equiv \int d\omega \, \omega^{2\tau+2}
K_0(\omega)K_1(\omega)= \frac{\sqrt{\pi}}{4}
\frac{\Gamma^2\left(\tau+1\right)
\Gamma\left(\tau+2\right)}{\Gamma\left(\tau+\frac{3}{2}\right)} \,.
\end{equation}
Finally, comparing with equation (\ref{w}) and using the relation
(\ref{relWT}) we obtain the Chern-Simons (CS) term contribution to
the antisymmetric structure functions
\beq
F_3^{CS}\left(x,q^2\right) =
\frac{1}{x}\left(\frac{\Lambda^2}{q^2}\right)^{\tau-1}{\cal{Q}}\frac{\pi^2
|c_i|^2 }{6\sqrt{4\pi\lambda}}{\cal{I}}_{\tau} \, , \label{F3}
\eeq
and $g_1^{CS}\left(x,q^2\right) = g_2^{CS}\left(x,q^2\right) = 0$.

Note that there is no contribution proportional to $S_\mu$ due to
the $R$-current conservation. This is similar to the case of the
symmetric part described in section 2. However, in the antisymmetric
part there is an important difference. There are examples of
holographic dual models similar to ${\cal{N}}=4$ SYM in the UV but
where the $R$-symmetry is spontaneously broken in the
IR\footnote{For a specific example see \cite{Hatta:2009ra} and
references therein.}. As noted in \cite{Hatta:2009ra}, in those
models our computation actually leads to a non-zero result for
$g_1$, more precisely we have
\begin{equation}
g_1^{CS}(x,q^2) = \frac{1}{2} \, F_3^{CS}(x,q^2) \propto \frac{1}{x}
\, . \label{2g1F3}
\end{equation}
With respect to $g_1^{CS}$, from now on we assume that we work with
a model of this kind. This will be important in order to analyze the
phenomenological implications of our results for the antisymmetric
contributions in section 5.

Let us briefly comment on the results we have obtained so far. We
have obtained new relations of the Callan-Gross type for the
antisymmetric structure functions in the range $\lambda^{-1/2} \gg x
\gg e^{-\sqrt{\lambda}}$. They can be compared for example with the
corresponding ones in the $1 > x \gg \lambda^{-1/2}$ region
\cite{Gao:2009ze}. We find that the relation $F_3 = 2 g_1$ holds for
low $x$. In addition, the structure function $g_2$ vanishes for
$\lambda^{-1/2} \gg x \gg e^{-\sqrt{\lambda}}$, which could seem to
be surprising since in the $x \sim 1$ regime $g_2$ is of the same
order as $F_3$, however, this is in agreement with the discussion
presented in reference \cite{Hatta:2009ra}. We should emphasize that
in the string theory regime the structure functions $F_3$ and $g_1$
are of the same order as $F_2$.  A similar behavior was found in the
scalar case for $F_3$ \cite{Kovensky:2017oqs}. Also, it is important
to note that, in contrast to the symmetric case, the antisymmetric
structure functions we have obtained are proportional to the
dilatino charge ${\cal {Q}}_C$.

~

%
{\bf Contribution of the Pauli term}
%

~

Up to this point we have not considered the Pauli interaction
arising from the Lagrangian (\ref{Lpau}). The computation is
analogous to the one made in the previous section. One arrives to an
$F_3^{P}(x,q^2)$ structure function that behaves in a similar way as
the one we found in equation (\ref{F3}) with two differences:
firstly instead of $d_{33C}{\cal{Q}}^C$ the new structure function
is proportional to $d_{33C} \beta^C$, and secondly there is an extra
twist-dependent factor $(\tau-1)$. The relation between the
Chern-Simons (CS) and the Pauli (P) contributions to the $g_1$
structure function can be summarized as
\bea
\frac{g_1^P}{d_{33C} \beta^C} \propto
\frac{g_1^{CS}}{d_{33C}{\cal{Q}}^C}(\tau-1) \, ,
\eea
and similarly for $F_3$, which means that the Pauli contribution
becomes more important for hadrons with larger twist.

It is interesting to remark that from a holographic dual string
theory perspective the mechanism which leads to the antisymmetric
structure functions at low $x$ is very different with respect the
one in the $1 > x \gg \lambda^{-1/2}$ range studied in
\cite{Gao:2009ze} and \cite{Gao:2010qk}. In the latter regime,
$F_3$, $g_1$ and $g_2$ (together with $g_3$, $g_4$ and $g_5$) come
from the right handed nature of the (massless) dilatino solution in
AdS$_5$ near the boundary. On the other hand, in the low-$x$ regime,
they come from the non-Abelian Chern-Simons and Pauli terms. As it
has been shown in \cite{Kovensky:2017oqs} this also happens in the
dilaton case.

~

The total contribution to the antisymmetric structure functions is
given by the sum to $F_3=F_3^{CS}+F_3^P$ and $g_1=g_1^{CS}+g_1^P$.
Note that both the Chern-Simons and Pauli contributions lead to the
same dependence on the Bjorken parameter as well as on the virtual
photon momentum for each antisymmetric function. Thus, for each
function the only difference is a multiplying constant which can be
fixed through an overall fitting. We shall do this in section 5.

%
\section{DIS from spin-$1/2$ hadrons at exponentially small $x$: the Regge region}
%

In this section we consider the parametric region where the Bjorken
parameter becomes exponentially small. In this regime the so-called
ultra-local approximation does not hold. In terms of the
four-dimensional center of mass energy and the 't Hooft coupling,
this occurs when both $s$ and $\lambda$ are large but satisfy the
relation
\begin{equation}
\frac{\log (s)}{\lambda^{1/2}} = {\mathrm{constant}} \, .
\end{equation}
We can understand this by looking at the factor
$m^{\al'\tilde{t}/2}$ in the imaginary part of the string theory
scattering amplitude pre-factor in equation (\ref{IMG}), which
within this parametric region cannot be replaced by a constant. The
point is that even if the four-dimensional Maldestam variable $t$
vanishes, the ten-dimensional one $\tilde{t}$ could be non-vanishing
\cite{Polchinski:2002jw}. Moreover, in terms of our field solutions
the radial component\footnote{Here we neglect the contributions from
the angular directions on the $S^5$. These differential operators
are present for $\tilde{s}$ as well, but they can be neglected in
comparison to the four-dimensional contribution proportional to
$s$.} corresponds to a Laplacian acting on the mode that propagates
in the $t$-channel.

Let us briefly review the general ideas used in this context. First,
it is useful to define a spin-$j$ second order differential operator
according to
\begin{equation}
\Delta_j = z^2 \der_z^2 + (2j-3)z\der_z+ j(j-4) \, .
\end{equation}
This operator can only differ from the actual Laplacian only by a
($j$-dependent) constant, {\it i.e.} $\nabla^2_j = \Delta_j + f(j)$.
These constants can be fixed by looking at the supergravity
equations of motion. The case $j=2$ corresponds to the propagation
of a graviton in the $t$-channel and it gives $f(2)=0$. On the other
hand, $f(1)=3$ corresponds to the propagation of a gauge field. In
terms of the scattering amplitude, we see that the introduction of
the differential operator effectively breaks the local
approximation. It leads to an effect of diffusion in the
$z$-direction, which modifies the amplitude dependence on $q^2$. As
we will see, it also gives an ${\cal{O}}\left(\lambda^{-1/2}\right)$
correction to the exponent of $1/x$ in the structure functions.
There are different ways of dealing with this operator, which are
described in detail in references \cite{Brower:2006ea,Brower:2007xg}
for $j\approx 2$, which is associated with a Reggeized graviton
exchange. This is the relevant case for the symmetric structure
functions. On the other hand, the spin-$j\approx 1$ exchange has
been studied in references \cite{Hatta:2009ra,Kovensky:2017oqs}, and
it is important for the antisymmetric structure functions. Let us
briefly explain how it works. We can assume that the operator
$\Delta_j$ acts on a generic field $\Phi(z)$. Then,
\bea
(\al'\tilde{s})^{\al' \tilde{t}/2} \Phi(z) &=&
(\al'\tilde{s})^{\rho\nabla_j^2/4} \Phi(z)
\label{Kdelta}\\
&=& \int dz' \, (\al'\tilde{s})^{\rho[\Delta_j+f(j)]/4} \delta(z-z') \Phi(z') \nn \\
&=& \int \frac{dz'}{z'}  \, \left(\frac{z'}{z}\right)^{j-2}
\frac{\left[e^{\zeta \rho}\right]^{\frac{1}{4}(f(j)-4)}}{\sqrt{\pi
\zeta \rho}} \, e^{-\frac{1}{\zeta \rho}\log^2(z/z')}\Phi(z') \, ,
\nn
\eea
where we have defined
\begin{equation}
\rho \equiv 2/\sqrt{\lambda} \ , \ \zeta \equiv \log{\al' \tilde{s}}
= \log \left(\al' zz' s\right) \, ,
\end{equation}
and inserted a Dirac delta function written in the form
\begin{equation}
z' \, \delta(z-z') = \int \frac{d\nu}{2\pi}
\left(\frac{z'}{z}\right)^{j-2+i\nu} = \int \frac{d\nu}{2\pi}
e^{(j-2+ i \nu) (u-u')}\,,
\end{equation}
where $z = e^{-u}$. This is equivalent to considering the identity
written in terms of eigenfunctions of the operator $\Delta_j$.
However, for non-conformal backgrounds one has to impose suitable
boundary conditions. In the hard-wall model where there is a cut-off
at $z=z_0$ we can impose Neumann-type conditions. Thus, only a
different linear combination of the previous eigenfunctions
survives. This is taken into account by the following replacement
\begin{equation}
e^{i \nu u } \rightarrow e^{i \nu u } + \left(\frac{\nu -2 i}{\nu +
2i}\right)e^{-i \nu u} \, .
\end{equation}
The only change in the final expression analogous to (\ref{Kdelta})
is
\begin{equation}
e^{-\frac{1}{\zeta \rho}\log^2(z/z')} \rightarrow e^{-\frac{1}{\zeta
\rho}\log^2(z/z')} + {\cal{F}}(z,z',\zeta) \, e^{-\frac{1}{\zeta
\rho}\log^2(z z'/z_0^2)} \, .
\end{equation}
The function $-1<{\cal{F}}(z,z',\zeta)<1$ is defined as in
\cite{Brower:2010wf} (equation (5.8) of that reference).

The expressions we have obtained are related to the so-called
Pomeron exchange. By inserting the particular case $j \approx 2$ and
$j \approx 1$ into the respective on-shell effective actions we
obtain the explicit integral form for the leading contributions to
the amplitude of the dual FCS process. Schematically, the general
amplitude takes the form
\begin{equation}
{\cal{A}} = 2 s \int d^2b \, \int dz dz' \, P_{A}(z)P_{\psi}(z') \,
\chi(z,z',s,b)|_{t=0} \, , \label{A1pomeron}
\end{equation}
in this kinematic regime where $P_A(z)$ contains the information on
the gauge field wave-function and $P_{\psi}(z)$ contains that of the
dilatino\footnote{In particular, in the present context it is
proportional to the generic $\Psi(z')$ defined above.}, while $\chi$
is the corresponding modified propagator. By taking the imaginary
part, this propagator corresponds to the conformal and the hard-wall
Pomeron kernels. We present the explicit forms of all these
quantities for the cases of interest in the following section.
Although in the DIS context we have been dealing only with the
imaginary part of ${\cal{A}}$, the expression in equation
(\ref{A1pomeron}) is valid for the real part as well. It has also
been extended to non-zero $t$, which corresponds to non-zero
transverse momentum transfer in the flat directions, and it has also
been written in the impact-parameter space. The form of these
amplitudes shows a formal similarity to the weak-coupling BFKL
results \cite{Brower:2006ea,Brower:2007xg}.

Under certain conditions, the analysis described in the previous
paragraphs has been extended in order to include an infinite number
of $1/N^2$ corrections. This has been achieved in the context of the
eikonal approximation, which gives the sum of all the $t$-channel
ladder diagrams shown in figure 5.
\begin{figure}[ht]
\centering
\includegraphics[scale=1]{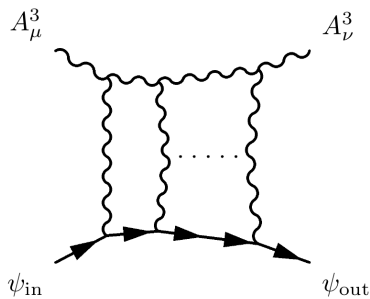}
\label{} \caption{\small Ladder diagrams re-summed by the eikonal
exponentiation, corresponding to a multi-Pomeron exchange.}
\end{figure}
The flat-space eikonal approximation is well-known, and its
extension to the curved AdS$_5$ background for the graviton and a
($j\approx 2$) Pomeron exchanges was investigated in several
important papers
\cite{Brower:2007qh,Brower:2010wf,Cornalba:2006xm,Cornalba:2007zb}.
In the eikonal regime, by including the higher orders in the ladder
expansion, it leads to an exponentiation of the amplitude of
equation (\ref{A1pomeron}) in the form
\begin{equation}
{\cal{A}} = 2 i s \int d^2b \, \int dz dz' \, P_{A}(z) \,
P_{\psi}(z') \, \left[1-{\mathrm{exp}}\left(i \,
\chi(z,z',s,b)|_{t=0}\right)\right] \, , \label{Aeikonal}
\end{equation}
where $b$ is the impact parameter. The previous case corresponds to
the first non-trivial term in the power series expansion of the
exponential. We should stress that in the present work we only
consider a single Pomeron exchange. The following contributions are
then associated with multi-Pomeron exchanges, each order being
suppressed by a $1/N^2$ factor. The implications of these results in
the context of the unitarity restrictions and the saturation regime
were studied in \cite{Hatta:2007he}.

\medskip

The kind of factors in equation (\ref{Kdelta}) present in the
amplitude are crucial when analyzing both the $x$ and $q^2$
dependence of the DIS amplitude and the structure functions. The
Bjorken parameter is contained in the $\zeta$-variable since for low
$x$ one has
\begin{equation}
\zeta = \log \left(\al' zz' s\right) \approx \log \left( z z'
\frac{\rho}{2} \frac{q^2}{x} \right) \, .
\end{equation}
Thus, we straightforwardly see that the behavior $1/x^2$ and $1/x$
of the different structure functions at small but not exponentially
small values of $x$ has to be modified by an extra factor
\begin{equation}
\left[e^{\zeta \rho}\right]^{\frac{1}{4}(f(j)-4)}\sim
\left(\frac{1}{x}\right)^{-\frac{1}{2\sqrt{\lambda}}(4-f(j))} \, .
\end{equation}
For $f(j)<4$, this implies that the increase of the amplitudes and
the structure functions when $x \rightarrow 0$ becomes softened. The
actual correction depends on the spin of the propagating field. For
the Reggeized graviton, $j\approx 2$ hence $f(j)\approx 0$ (up to
${\cal {O}}(\lambda^{-1/2})$ terms), leading to a correction in the
exponent of $-\frac{2}{\sqrt{\lambda}}$. This correction affects all
the functions contained in the symmetric part of the hadronic
tensor. On the other hand, when the exchanged field is a Reggeized
gauge field $j\approx 1$, which implies that $f(j)\approx 3$. Thus,
the correction is somewhat less important, namely:
$-\frac{1}{2\sqrt{\lambda}}$. Therefore we conclude that, for
exponentially small $x$ as $x \rightarrow 0$, the antisymmetric
structure functions obtained in this work grow faster than $F_2$ in
the same parametric regime.

Having described the general considerations, let us write down the
explicit expressions for the structure functions at tree-level and
in the hard-wall model. Due to the energy-momentum tensor
conservation we have seen that for small values of $x$ the symmetric
part of the hadronic tensor is dominated by the universal
contributions associated with $F_1$ and $F_2$. Inserting the
corresponding (imaginary part of) the $j\approx 2$ kernel one
obtains the following expression for the $F_2$ \cite{Brower:2010wf}
\begin{equation}
F_2(x,q^2) \sim
\int \frac{dz}{z} \frac{dz'}{z'} P_{A}(z,q^2) P_{\psi}(z')
(zz'q^2)\,
\frac{e^{\zeta(1-\rho)}}{\sqrt{\zeta}}\left(e^{-\frac{\log^2\left(z/z'\right)}{\rho
\zeta}} +{\cal{F}}(z,z',\zeta)
e^{-\frac{\log^2\left(zz'/z_0^2\right)}{\rho \zeta}} \right) \, .
\label{F2brower}
\end{equation}
The wave-function dependent terms are defined as
\begin{equation}
P_{A}(z,q^2) = (q z)^2 \left(K_1^2(q z) + K_0^2(q z)\right) \ , \
P_{\psi}(z') = z^{-3} |f_+(z')|^2 \sim (z' \Lambda)^{2\tau-2} \, ,
\end{equation}
being $f_+(z')$ given in terms of the initial state $\psi(x,z) =
e^{i P\cdot x} \left[f_{+}(z) P_+ + f_{-}(z) P_- \right]u(P)$ as
equation (\ref{PsiaaaC}). In the last formula we have written its
near-boundary expansion. The analogous expression for $2x
F_1(x,q^2)$ can be obtained by omitting the $K_0^2$ term in $P_{A}$.

For the case of the antisymmetric contributions one has to use the
$j\approx 1$ kernel. In this context it is interesting to consider
situations in which the $R$-symmetry is spontaneously broken in the
IR in such a way that it allows for a contribution to the
$g_1(x,q^2)$ structure function. In such case, $g_1$ does not
vanish, moreover it is related to $F_3$ through  $2 g_1 = F_3$.
Assuming that the kernels can be approximately described in the same
way, this leads to the holographic dual description of the structure
function $g_1$ at low $x$:
\bea
g_1(x,q^2) &=& \frac{{\cal{Q}}\pi^2}{24} \int dy\,dy' \,
{\cal{P}}_A(y,q) P_\psi (y') \, \times \nn \\
&&  \,
\frac{e^{\zeta \left(1-\rho/4\right)}}{\sqrt{\pi \zeta \rho}} \left(
e^{-\frac{\sqrt{\lambda}}{8\zeta}(y-y')^2} + {\cal{F}}(y,y',\zeta)
e^{-\frac{\sqrt{\lambda}}{8\zeta}(y+y')^2} \right)
   \label{F3pomeron} \, ,
\eea
with $y=-2\log(z)$. Note that the $P_A$ factor has to be replaced by
\begin{equation}
{\cal{P}}_{A}(z,q^2) = (q z)^3 K_1(q z) K_0(q z) \, ,
\end{equation}
reflecting the use of the Chern-Simons term. For more details of the
computation we refer the interested reader to references
\cite{Hatta:2009ra,Kovensky:2017oqs}. Equation (\ref{F3pomeron}) is
very similar to the one obtained for $F_3$ in the scalar case in
\cite{Kovensky:2017oqs}.

Equations (\ref{F2brower}) and (\ref{F3pomeron}) are difficult to
evaluate analytically.  In the next section we describe a possible
way to extract the relevant information, from which we perform a
very interesting phenomenological analysis.

%
\section{Analysis of the results and conclusions}
%

In reference \cite{Polchinski:2002jw} Polchinski and Strassler have
shown that in the region $1 > x \gg \lambda^{-1/2}$ the results
obtained from the computation of the $s$-channel amplitude in the
supergravity approximation are different from the QCD expectations
in the parton model for weak coupling. This is partly because in the
planar limit, where the supergravity approximation holds, particle
creation in the bulk becomes suppressed. It means that the virtual
photon interacts with the entire hadron since the latter does not
effectively contain partons in that limit. The structure functions
show a behavior of the form $(\Lambda^2/q^2)^{\tau-1}$. This is
related to the fact that the hadron wave-function is localized near
$z_0=\Lambda^{-1}$. Thus, in order for inelastic scattering to
occur, the string (which holographically represents the hadron) must
tunnel to the region near the boundary ($z < q^{-1}$).

Let us very briefly consider the $x$-dependence of the structure
functions in the region $1 > x \gg \lambda^{-1/2}$. They are somehow
similar to bell-shaped curves, with maxima around $ x_{\star}\sim
0.6$ which are larger than the experimental observations. In this
regime, these results have been extended for charged and neutral
polarized spin-$1/2$ hadrons \cite{Gao:2009ze,Gao:2010qk}, and also
for scalar and polarized vectors mesons for different Dp-brane
models, both in the Abelian and non-Abelian cases
\cite{Koile:2011aa,Koile:2013hba}. However, by considering
supergravity one-loop corrections in this $x$-regime very
interesting results have been found
\cite{Jorrin:2016rbx,Kovensky:2016ryy}. Particularly, the first
moments of the structure function $F_2$ for the pion
\cite{Koile:2011aa,Koile:2013hba} can be compared with the lattice
QCD ones \cite{Best:1997qp}, having found discrepancies under
20$\%$, while the supergravity one-loop level calculation improves
quite significantly that prediction reaching an accuracy of 1.27
$\%$ or better. Similarly, it occurs for the first three moments of
$F_1$ of the rho meson, which gives tree-level results with accuracy
of 20$\%$ \cite{Koile:2011aa,Koile:2013hba}, while for the one-loop
level calculations the agreement with respect to lattice QCD results
\cite{Best:1997qp} reaches an accuracy of 3$\%$
\cite{to-appear-2018}\footnote{It is interesting to also compare the
level of accuracy for other physical observables calculated in terms
of the AdS/CFT duality. For instance in the bottom-up AdS/QCD model
observables obtained by the calculation of two-point functions lead
to an overall fitting of 5$\%$ or better
\cite{Erlich:2005qh,DaRold:2005mxj}. For observables depending on
four-point functions, for instance for the $\Delta I=1/2$ rule for
the kaon decay the level of agreement is about 25$\%$ or better
\cite{Hambye:2005up,Hambye:2006av}.}.

On the other hand, at low $x$ the holographic dual calculation must
include the exchange of excited string states in the $t$-channel.
This approach leads to important new insights on Regge physics since
the AdS/CFT duality provides a unified description for both the soft
Pomeron and the BFKL Pomeron \cite{Brower:2006ea}. In QCD when $q^2$
becomes small, it is not possible to think of the constituents of a
hadron as approximately free partons. Confinement as well as
saturation effects become important. These phenomena are related to
modifications of the Pomeron kernel and the inclusion of
multi-Pomeron exchange respectively. In reference
\cite{Brower:2010wf} it has been done a remarkable comparison with
$F_2(x,q^2)$ data for the proton obtained by the HERA Collaboration
\cite{Aaron:2009aa,Breitweg:1998dz,Chekanov:2001qu}.

Also the gauge/string theory duality has been applied to other
scattering processes such as deeply virtual Compton scattering
(DVCS), double diffractive Higgs production, generalized parton
distributions (GPD) and form factors\footnote{See for example
\cite{Nishio:2011xz,Costa:2012fw,Watanabe:2012uc,Costa:2013uia,Nally:2017nsp}
and references therein.}.

%
\subsection{Structure functions results at low $x$}
%

In the first part of this work we have computed both the symmetric
and antisymmetric structure functions of a polarized spin-1/2 hadron
in the low $x$ regime. The target is represented by a dilatino KK
mode in the dual type IIB superstring theory on AdS$_5\times S^5$,
with an IR deformation. This has been done in two separate but
equivalent ways: from superstring theory scattering amplitudes and
in a heuristic way developed from the supergravity interactions.

For the symmetric structure functions we have shown that $F_1$ and
$F_2$ behave as $x^{-2}$ and $x^{-1}$, respectively, and found a new
generalized Callan-Gross relation given by equation (\ref{F222}). It
is analogous to the one found for the scalar glueball
\cite{Polchinski:2002jw} and also in the meson case
\cite{Koile:2014vca}. Moreover, the $g_{3,4,5}$ vanish in this
regime, in contrast to the results obtained in the $1>x\gg
\lambda^{-1/2}$ regime \cite{Gao:2009ze}. Although it is not obvious
at first sight from the gravity computation, from the CFT point of
view we know that the dominant contribution to the $JJ$ OPE is
proportional to the energy-momentum tensor. Since the
$g_{3,4,5}$-terms in the hadronic tensor are proportional to the
spin vector $S^\m$, they cannot appear at this order.

By considering the $t$-channel $j\approx 1$ exchange, we have also
described the leading contributions to the antisymmetric part of the
hadronic tensor. This leads to $g_1 \sim x^{-1}$ (and the same for
$F_3$). Strictly speaking, an argument similar to that of the
previous paragraph implies that in the hard-wall model $g_1$ should
vanish. However, there is an important difference: here the relevant
term comes from the $JJ \sim J$ term in the OPE. If one considers a
QFT where the $R$-symmetry is spontaneously broken in the IR, it is
possible to obtain a non-vanishing $g_1 = \frac{1}{2} F_3$. On the
other hand we find that the structure function $g_2$ vanishes in
this regime. Recall that in the parton model $g_2$ also vanishes,
moreover there is no simple interpretation for this function in the
parton model \cite{Manohar:1992}. In the following we shall
concentrate on the comparison of our results for the
phenomenologically relevant structure function $g_1$ with respect to
experimental data.

%
\subsection{New predictions for $g_1$ and comparison with COMPASS data}
%

In the hard-wall model, symmetric structure functions at low-$x$ and
exponentially small-$x$ depend on a finite set of parameters as
shown in equation (\ref{F2brower}). Since the $z$ and $z'$ integrals
are difficult solved analytically, it has been proposed to
approximate the $P_{i}$ factors by Dirac delta functions supported
at appropriate reference scales \cite{Brower:2010wf}
\bea
P_{\psi}(z')\approx \frac{1}{q'} \delta(z'-1/q') \, , \,\,\,\,\,\, P_{A}(z)
\approx \frac{1}{q}\delta(z-1/q) \, , \label{deltas}
\eea
where $q'$ is some scale of order of the hadron mass. Using this
approximation the integrals in equation (\ref{F2brower}) can be
performed, obtaining an expression that has four free parameters,
namely: an overall constant, $z_0$, $\rho$ and $q'$. These
parameters can be fixed by fitting $F_2$ to data. In fact, there is
considerable amount of experimental results from HERA obtained from
DIS off protons at low $x$ which have been used in order to fit the
structure function $F_2$. This has been done by considering the
H1-ZEUS data \cite{Aaron:2009aa,Chekanov:2001qu,Breitweg:1998dz} for
$x<10^{-2}$. In \cite{Brower:2010wf} the authors fitted $F_2$ for
the conformal and the hard-wall models. In the conformal model their
fit leads to
\begin{equation}
\rho = 0.7740\pm 0.0103 \ , \ q' = 0.5575 \pm 0.0432\,
{\mathrm{GeV}}\ , \label{parametros_conf}
\end{equation}
with a reduced chi-square $\chi_{d.o.f.}^2 = 0.75$. On the other
hand, their best fit for $F_2$ using the hard-wall model gives
\cite{Brower:2010wf}
\begin{equation}
\rho = 0.7792\pm 0.0034 \ , \ q' = 0.4333 \pm 0.0243\,
{\mathrm{GeV}}\ , \ z_0 = 4.96 \pm 0.14\, {\mathrm{GeV}}^{-1}\, ,
\label{parametros_HW}
\end{equation}
with $\chi_{d.o.f.}^2 = 1.07$. Interestingly, their values of the
parameters are reasonable since $\rho$ lies between the strong/weak
coupling transition, $q'$ is of order of the proton mass, while $z_0
\sim {\cal {O}}(\Lambda_{\mathrm{QCD}}^{-1})$ as expected in
hard-wall model.

Now, we can obtain new predictions for the $g_1$ structure function
by using the formal expressions derived in our present work.
Considering the approximation (\ref{deltas}) in equation
(\ref{F3pomeron}) we obtain the following expression for the
antisymmetric structure function $g_1$ which holds for both the
conformal (${\cal{F}}=0$) and the hard-wall kernels
\bea
g_1(x,q^2) &=& {\cal{C}}\, \rho^{-1/2} e^{\zeta(1-\rho/4)}
\left\{\frac{\exp \left[ -\frac{\log^2 (q/q')}{\rho \zeta}
\right]}{\sqrt{\zeta}}+{\cal{F}}(q,q',\zeta) \frac{\exp \left[
-\frac{\log^2 (q q' z_0^2)}{\rho \zeta} \right]}{\sqrt{\zeta}}
\right\} \nonumber \\
&\approx & {\cal{C}}\, \frac{\rho^{1/2}}{2x}
\frac{q}{q'}e^{\zeta(-\rho/4)} \left\{\frac{\exp \left[
-\frac{\log^2 (q/q')}{\rho \zeta}
\right]}{\sqrt{\zeta}}+{\cal{F}}(q,q',\zeta) \frac{\exp \left[
-\frac{\log^2 (q q' z_0^2)}{\rho \zeta} \right]}{\sqrt{\zeta}}
\right\} \, , \label{g1fenom}
\eea
where we have used that $e^{\zeta} \approx
\frac{1}{\sqrt{\lambda}}\frac{1}{x} \frac{q}{q'}$.
Next, we carry out the comparison with experimental data. For that,
firstly note that three of the four free parameters in equation
(\ref{g1fenom}) are already determined by the previous fitting of
the structure function $F_2$ done in \cite{Brower:2010wf} and shown
in equation (\ref{parametros_HW}) (or (\ref{parametros_conf}) for
the conformal model). Therefore, the only unknown free parameter is
the overall constant ${\cal{C}}$.

\begin{figure}[h!]
\centering
\includegraphics[scale=1]{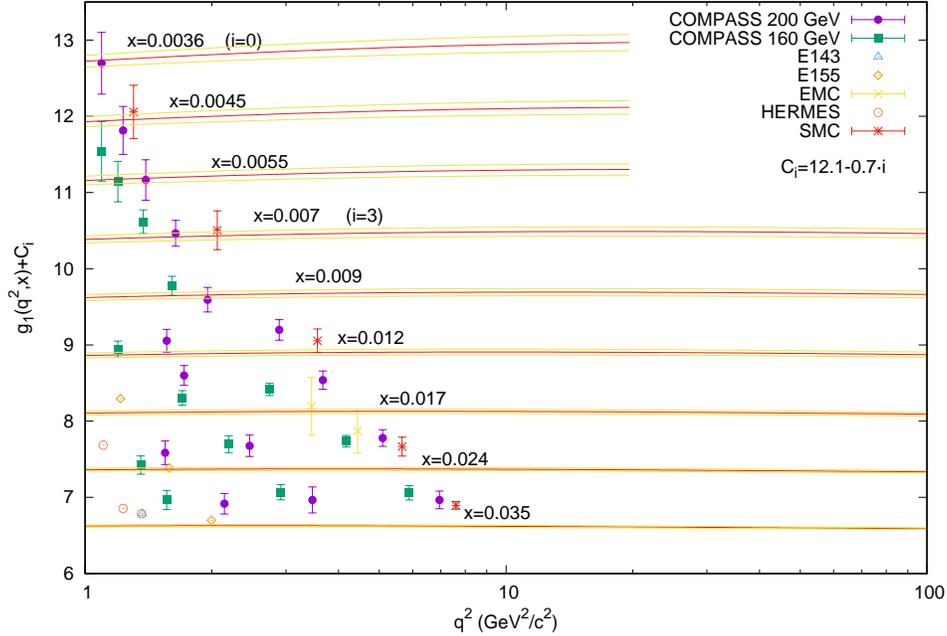}
\label{} \caption{\small  Red curves display our fit for the $g_1$
structure function as a function of $q^2$ for different values of
the Bjorken variable, compared to the experimental data presented in
\cite{Adolph:2015saz} and references therein, while the yellow lines
are the values of $g_1$ considering the error on ${\cal{C}}$. The
best fit corresponds to $x<0.01$ (larger values of the Bjorken
parameter are shown for completeness) obtaining a constant
${\cal{C}} = 0.0195 \pm 0.0024$ with a $\chi_{d.o.f.}^2 = 0.27$.
Note that following reference \cite{Adolph:2015saz} for each value
of $x$ we are adding a constant $C_i = 12.1 - 0.7 i$ to the $g_1$
data and the corresponding curve.}
\end{figure}

\begin{figure}[h!]
\centering
\includegraphics[scale=1]{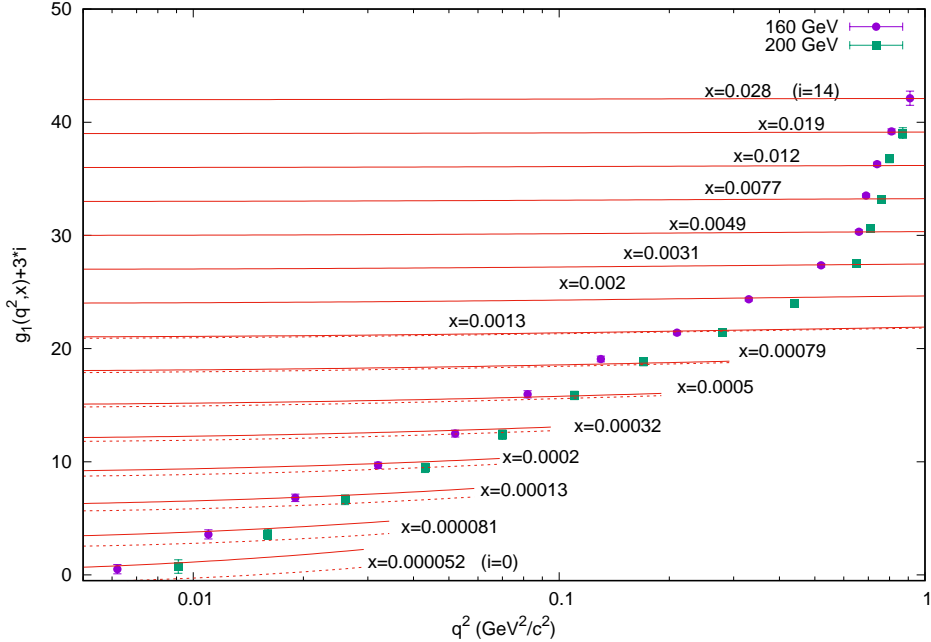}
\label{} \caption{\small  Our best fit of the structure function
$g_1$ carried out with the newest data presented in
\cite{Aghasyan:2017vck}. Solid curves correspond to the best fit for
the conformal model, with ${\cal{C}} = 0.0112 \pm 0.0020$ and a
$\chi_{d.o.f.}^2 = 1.140$, while dotted lines correspond to the best
fit for the hard-wall model with ${\cal{C}} = 0.0120 \pm 0.0020$ and
a $\chi_{d.o.f.}^2 = 1.074$. For values above $x=0.0013$ we only
show the conformal model because at the scale used in the plot there
are no visible differences. In this case we do not show the error on
${\cal{C}}$ because it is negligible in the displayed scale. Note
that for each row corresponding to different $x$ value, we are
adding the constant $3 i$.}
\end{figure}

The $g_1$ structure function of the proton has been measured by the
SMC Collaboration \cite{Adeva:1998vv}, also more recently by the
COMPASS Collaboration with the beam energies of $160$ GeV and $200$
GeV \cite{Alekseev:2010hc,Adolph:2015saz}. The corresponding sets of
data can be found in the mentioned experimental references. Since
the calculations performed in the previous sections are valid for
the low-$x$ regime, we consider data within the $x<0.01$ region.
Therefore, we have nineteen experimental values that can be used to
fit equation (\ref{g1fenom}). Proceeding in this way we obtain the
constant value: ${\cal{C}} = 0.0195 \pm 0.0024$ for the conformal
model, and ${\cal{C}} = 0.0191 \pm 0.0023$ for the hard-wall model.
Both fits give $\chi_{d.o.f.}^2 = 0.27$, which indicates that our
model is over-fitting the data set, thus this is not a good fit. In
figure 6 the experimental data presented in \cite{Adolph:2015saz}
and our first fit for the conformal model are shown. For
completeness we included experimental points up to $x=0.035$
obtained by the SMC \cite{Adeva:1998vv}, EMC \cite{Ashman:1987hv},
HERMES \cite{Airapetian:2006vy}, SLAC E143 \cite{Abe:1998wq}, E155
\cite{Anthony:2000fn} and CLAS \cite{Prok:2014ltt} collaborations,
at $q^2 > 1$ (GeV/c)$^2$. The fit is better in the region $x<0.01$
and then it drifts apart from the experimental data as $x$
increases, {\it i.e.} where the Pomeron approach does not hold.

It is very interesting to compare with the most recent data from the
COMPASS Collaboration. They have published new and more precise data
for the $g_1$ structure function of the proton
\cite{Aghasyan:2017vck} for photon virtuality $q^2 < 1$ (GeV/c)$^2$
and for the Bjorken parameter $4 \times 10^{-5}  < x < 4 \times
10^{-2}$. This region seems more suitable for our analysis given
that the Pomeron formalism describes DIS processes at very low
values of $x$. However, one should keep in mind that we have
considered $q$ to be much larger than the $\Lambda$ scale. Thus, it
seems reasonable to consider only the data set where $q^2 > q'^2$,
being the latter approximately $0.2$-$0.3$ (GeV/c)$^2$ according to
\eqref{parametros_conf} and \eqref{parametros_HW}.
In this way, we consider fifteen points, which represent half of the
data presented in \cite{Aghasyan:2017vck}, and obtain the following
results for the fits:
\bea
{\mathrm{conformal \,\, model}}&:& \ {\cal{C}} = 0.011
\pm 0.002\ ,\, \chi_{d.o.f.}^2 = 1.140 \label{C-conforme} \\
{\mathrm{hard-wall \,\, model}}&:& \ {\cal{C}} = 0.012 \pm 0.002\
,\, \chi_{d.o.f.}^2 = 1.074  \,. \label{C-HW}
\eea
As we can see, for this new data set we obtain a very good fit. The
value of the parameter ${\cal{C}}$ does not change significantly,
being always ${\cal{C}}\simeq 0.01$. This is a very interesting
prediction for the proton structure function $g_1$, and together
with expression \eqref{g1fenom} it represents the main result of
this work.
As expected, the confining model gives a more accurate description
in the region where $q$ and $q'$ become comparable
\cite{Brower:2010wf}.
As an aside, we should say that, rather surprisingly, including all
data points from \cite{Aghasyan:2017vck} it still renders an
acceptable fit (with $\chi_{d.o.f.}^2 = 0.911$ and ${\cal{C}} =
0.0114 \pm 0.0011$), but only for the conformal model. This is not
so in the confining case.
Figure 7 displays the experimental data presented in
\cite{Aghasyan:2017vck} together with our best fits.
It can be seen that the hard-wall model gives the best fit for the
points with larger values of $q^2$, while the conformal model gives
an acceptable fit for the full data set.
Finally, by using our results \eqref{C-conforme} and \eqref{C-HW} we
can predict the behavior of $g_1(x, q^2)$ for different values $q^2$
in the small-$x$ regime. This is shown in figure 8.

In order to make the comparison between our fitted $g_1$ function
and the experimental data simpler, figures 6 and 7 are displayed in
a similar way as the corresponding figures of references
\cite{Adolph:2015saz} and \cite{Aghasyan:2017vck}, respectively. We
briefly comment on the range of validity of the fits we have done.
The first set of data described in figure 6 corresponds to larger
values of $x$ and a relatively broad range of $q^2$. On the other
hand, the new set of data shown in figure 7 corresponds to much
lower values of the Bjorken parameter, although the $q^2$-range is
much smaller (notice the logarithmic scale for $q^2$ in both
figures). The ideal situation where our results should fit better
data would be for low $x$ and large photon virtuality. It would be
very interesting to have experimental data in that parametric range.
We should also emphasize that the amount of experimental data for
$g_1$ at present is much less than the available data for $F_2$.
Thus, our predictions for $g_1$ in terms of the comparison with
experimental data could possibly be improved, depending on the
availability of new data in future.

\bigskip

\begin{figure}[h!]
\centering
\includegraphics[scale=1]{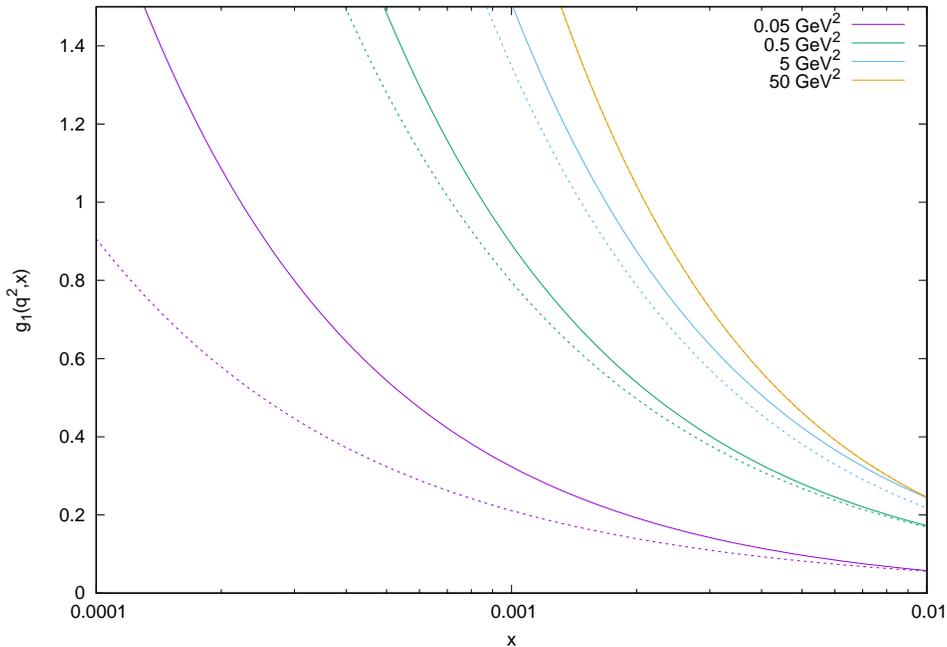}
\label{} \caption{\small Our results for the structure function
$g_1$. Solid lines correspond to the conformal model. Dotted lines
correspond to the hard-wall model. The values of the photon
virtuality $q^2$ are indicated.}
\end{figure}

There is an important and very interesting aspect to emphasize. We
know that in QCD the electromagnetic DIS leads to a vanishing $F_3$
structure function. This changes drastically when considering DIS
for weak interactions mediated by $W^{\pm}$ or $Z^0$ gauge bosons.
Thus we should stress that although QCD and this IR-deformed
${\cal{N}}=4$ SYM theory we consider can have a number of analogous
properties in the planar limit, as QFTs they are different. One
important difference is that ${\cal{N}}=4$ SYM theory is chiral. The
$R$-symmetry current associated with the global $U(1)_R \subset
SU(4)_R$ is promoted to become a gauge symmetry in order to describe
the electric current. Therefore, our prediction for a non-vanishing
$F_3$ is entirely related to an IR deformation of ${\cal{N}}=4$ SYM
theory.

On the other hand, we conclude that the present results for $g_1$
fit very well the experimental data as shown in figure 7. Let us
emphasize that our knowledge this is the first fully
string-theoretical derivation and comparison with experimental data
from a calculation of $g_1$ obtained by using the gauge/string
theory duality framework, where non-perturbative physics plays a
major role. Also, it is important to remark that the fits for $g_1$
presented above are totally compatible with those of $F_2$ obtained
by Brower et al \cite{Brower:2010wf}. Therefore, this work also
contributes to the understanding of a unified picture of the Pomeron
physics by adding relevant results from the holographic dual
description of the antisymmetric structure functions.

~

~

%
\centerline{\large{\bf Acknowledgments}}
%

~

We thank Edmund Iancu, Gino Marceca and Carlos N\'u\~nez for useful
discussions and comments. N.K. acknowledges kind hospitality at the
Institut de Physique Th\'eorique, CEA Saclay, and at the Institute
for Theoretical Physics, University of Amsterdam, during the
completion of this work. G.M. acknowledges kind hospitality at the
Instituto Balseiro, Bariloche, and at the International Center for
Theoretical Physics, Trieste, where part of this work has been done.
This work has been supported by the National Scientific Research
Council of Argentina (CONICET), the National Agency for the
Promotion of Science and Technology of Argentina (ANPCyT-FONCyT)
Grants PICT-2015-1525 and PICT-2017-1647, and the CONICET Grant
PIP-UE B\'usqueda de nueva f\'{\i}sica.

\newpage

%

\end{document}